\documentclass[10pt,times,a4paper]{iopart}

\expandafter\let\csname equation*\endcsname\relax
\expandafter\let\csname endequation*\endcsname\relax

\usepackage{times}
\usepackage[usenames]{color}
\usepackage{amsfonts}
\usepackage{amsmath}
\usepackage{amssymb}
\usepackage{bm}
\usepackage{dcolumn}
\usepackage{enumerate}
\usepackage{epsfig}
\usepackage{graphicx}
\usepackage{graphics}
\usepackage{multirow}
\usepackage[latin1]{inputenc}
\usepackage{latexsym}
\usepackage{rotating}
\usepackage{hyperref}
\usepackage[caption=false]{subfig}
\usepackage{float}
\usepackage[dvipsnames]{xcolor}
\usepackage{cite}

\newcommand{\<}{\begin{equation}}
\newcommand{\?}{\end{equation}}
\newcommand{\blambda}{\boldsymbol{\lambda}}

\newcommand{\dl}{d_{\text{L}}}
\newcommand{\mf}{M_{\text{f}}}
\newcommand{\af}{\chi_{\text{f}}}
\newcommand{\dmfbymf}{\frac{\Delta M_{\text{f}}}{\bar{M}_{\text{f}}}}
\newcommand{\dafbyaf}{\frac{\Delta \chi_{\text{f}}}{\bar{\chi}_{\text{f}}}}

\begin{document}

\title[Meta IMR consistency test of GR]{A meta inspiral-merger-ringdown consistency test of general relativity with gravitational wave signals from compact binaries}

\author{Sakshi Satish Madekar\footnote[1]{Author to whom any correspondence should be addressed.}}
\address{Department of Physical Sciences, Indian Institute of Science Education \& Research (IISER) Mohali, Sector 81 SAS Nagar, Manauli PO 140306, Punjab, India}
\ead{sakshi.0407.madekar@gmail.com}
\author{Nathan~K~Johnson-McDaniel}
\address{Department of Physics and Astronomy, University of Mississippi, University, Mississippi 38677, USA}
\ead{nkjm.physics@gmail.com}
\author{Anuradha Gupta}
\address{Department of Physics and Astronomy, University of Mississippi, University, Mississippi 38677, USA}
\ead{agupta1@olemiss.edu}
\author{Abhirup Ghosh}
\address{Trade Republic Bank GmbH, Brunnenstra{\ss}e 19-21, 10119 Berlin, Germany}
\ead{abhirup.ghosh.184098@gmail.com}

\date{\today}

\begin{abstract}
The observation of gravitational waves from compact binary coalescences is a promising tool to test the validity of general relativity (GR) in a highly dynamical strong-field regime. There are now a variety of tests of GR performed on the observed compact binary signals. In this paper, we propose a new test of GR that compares the results of these individual tests. This meta inspiral-merger-ringdown consistency test (IMRCT) involves inferring the final mass and spin of the remnant black hole obtained from the analyses of two different tests of GR and checking for consistency. If there is a deviation from GR, we expect that different tests of GR will recover different values for the final mass and spin, in general. We check the performance of the meta IMRCT using a standard set of null tests used in various gravitational-wave analyses: the original IMRCT, the Test Infrastructure for General Relativity (TIGER), the Flexible-Theory-Independent (FTI) test, and the modified dispersion test. However, the meta IMRCT is applicable to any tests of GR that infer the initial masses and spins or the final mass and spin, including ones that are applied to binary neutron star or neutron star--black hole signals. We apply the meta IMRCT to simulated quasi-circular GR and non-GR binary black hole (BBH) signals as well as to eccentric BBH signals in GR (analyzed with quasicircular waveforms). We find that the meta IMRCT gives consistency with GR for the quasi-circular GR signals and picks up a deviation from GR in the other cases, as do other tests. In some cases, the meta IMRCT finds a significant GR deviation for a given pair of tests (and specific testing parameters) while the individual tests do not, showing that it is more sensitive than the individual tests to certain types of deviations. In addition, we also apply this test to a few selected real compact binary signals and find them consistent with GR.
\end{abstract}

\section{Introduction}

Gravitational wave observations now regularly provide us with direct information about the highly dynamical strong-field mergers of binaries of black holes and neutron stars. The most recent completed observing run (O3) has brought the total number of detection candidates to around $100$~\cite{GWTC-3_paper}, and the ongoing observing run (O4, with some initial results released in~\cite{GW230529}) will considerably increase this number. There is now a considerable variety of tests of general relativity (GR) applied to the most confident detections---see~\cite{GW170817_TGR,O2_TGR,O3a_TGR,O3b_TGR} for work by the LIGO-Virgo(-KAGRA) (LVK) collaboration \footnote{We refer to the LV(K) collaboration for cases where the KAGRA collaboration only contributed to some of the results being referenced.} and~\cite{Johnson-McDaniel:2021yge} for an overview of some of the other tests being applied or proposed. However, all the tests currently applied are null tests of some type---none of them is testing a specific alternative theory. There are now some initial studies of the relation between a selection of these tests, looking at how they respond to various phenomenological non-GR signals~\cite{Johnson-McDaniel:2021yge} or signals beyond the standard quasicircular binary hypothesis used in the application of the tests~\cite{Narayan:2023vhm}, as well as strongly lensed signals~\cite{Narayan:2024rat} and signals from binaries with surrounding matter~\cite{Roy:2024rhe}. However, there is still much to be done to understand the extent to which these tests are complementary or redundant.

Here we propose a new method for comparing tests of GR, which itself provides yet another test of GR. Specifically, we generalize the inspiral-merger-ringdown (IMR) consistency test for binary black hole (BBH) signals introduced in~\cite{Ghosh:2016qgn,Ghosh:2017gfp} (which checks the consistency of the low- and high-frequency portions of the signal) to compare the posterior probability distributions for the mass and spin of the final black hole from two different tests of GR. As noticed in~\cite{Johnson-McDaniel:2021yge}, these distributions are often quite different when different tests of GR (including different versions of the same test, e.g.\ parameterized tests varying different post-Newtonian parameters) are applied to the same non-GR signal, or when comparing the results of a test of GR and a standard GR analysis of the signal. In these analyses, the final mass and spin are calculated by applying fits to BBH numerical relativity simulations that give the final mass and spin in terms of the binary's initial masses and spins. Thus, since the binary's initial masses and spins are biased in different ways for different analyses when attempting to reproduce a non-GR signal, the final mass and spin are also in general biased differently in different analyses. This is why this method can detect deviations from GR (or, more generally, deviations from the signal model used in the analysis, which is a quasicircular binary in all LVK tests of GR).


We illustrate the method, which we refer to as the \emph{meta IMR consistency test} (or meta IMRCT), using the results from the tests of GR (and standard GR parameter estimation analysis) applied to quasicircular GR and non-GR signals from~\cite{Johnson-McDaniel:2021yge} and quasicircular and eccentric numerical relativity signals from~\cite{Narayan:2023vhm} (here only analyzed with quasicircular waveform models). Specifically, we consider the TIGER~\cite{Meidam:2017dgf} and FTI~\cite{Mehta:2022pcn} parameterized tests, the modified dispersion relation test~\cite{Mirshekari:2011yq,O2_TGR}, and the standard IMR consistency test (IMRCT)~\cite{Ghosh:2016qgn,Ghosh:2017gfp}. Like the individual tests, the meta IMRCT finds quasicircular GR signals in~\cite{Johnson-McDaniel:2021yge,Narayan:2023vhm} consistent with GR but reports GR violation for non-GR and eccentric signals. We also apply the meta IMRCT to the compact binary signals GW170817~\cite{GW170817}, GW190412~\cite{GW190412}, GW190521~\cite{GW190521}, GW190814~\cite{GW190814} and GW200225\_060421~\cite{GWTC-3_paper}, using the results from the LV(K) collaboration's application of the aforementioned tests to these events. The meta IMRCT finds consistency with GR in these cases, as do the individual tests.

The paper is structured as follows: We introduce the existing tests of GR we consider and the Bayesian inference they are based on in section~\ref{sec:tests}, and then provide the specifics of the meta IMRCT in section~\ref{sec:meta_imrct}. We then give an overview of the simulated observations used to assess the performance of the tests in section~\ref{sec:sim_obs} and the results of the meta IMRCT in section~\ref{sec:results}. Finally, we conclude in section~\ref{sec:concl}. We use geometrised units with $G = c = 1$ throughout.

\section{Tests of GR}
\label{sec:tests}

This section provides an overview of standard GR parameter inference and the tests of GR that the meta IMRCT builds on. The original IMRCT~\cite{Ghosh:2016qgn,Ghosh:2017gfp}, the TIGER~\cite{Meidam:2017dgf} and FTI~\cite{Mehta:2022pcn} parameterised tests and the modified dispersion relation test~\cite{Mirshekari:2011yq,O2_TGR}, are now a staple of testing-GR analyses by the LVK and others. They have thus been described in detail in previous publications, e.g.~\cite{O2_TGR,O3a_TGR,O3b_TGR,Johnson-McDaniel:2021yge}. However, given the importance of the details of these analysis in motivating the meta IMRCT, we provide a brief review here again.

\subsection{Bayesian parameter estimation}

Before we review the tests, it is worth reviewing our parameter inference framework, which is Bayesian, as is the case for most of gravitational-wave data analysis. Data from a gravitational-wave detector is a noisy time series $\textbf{d}(t)$ which occasionally contains a gravitational-wave signal $\textbf{h}(t, \blambda )$, where $\blambda$ denotes the parameters describing the gravitational-wave signal. The data can thus be written as
\begin{equation}
    \textbf{d}(t) = \textbf{h}(t, \blambda) + \textbf{n}(t),
\end{equation}
where $\textbf{n}(t)$, the noise in detector data, is assumed to be Gaussian and characterised by a (frequency-domain) power spectral density $S_n(f)$. The gravitational-wave signal is described by a hypothesis $H$, which can either be GR or an alternate theory of gravity. Even within GR, waveforms of signals can be built with varying assumptions, and hence our hypothesis is the gravitational waveform used in the data analysis. This waveform is usually characterised by a multi-dimensional parameter set $\blambda$, which includes a GR parameter set $\blambda _ {\text{GR}}$ and, in the case of a non-GR waveform, any additional non-GR parameter(s) $\blambda _ {\text{nGR}}$ that describe it.

Given a hypothesis $H$ and any other information $I$, a posterior probability distribution of $\blambda$, $P(\blambda | \textbf{d}, I )$, can be inferred using the Bayes Theorem:
\begin{equation}
    P(\blambda | \textbf{d}, I ) \propto p(\blambda) \mathcal{L}( \textbf{d}| \blambda, I),
\end{equation}
where $p(\blambda)$ is the prior probability distribution and $\mathcal{L}( \textbf{d}| \blambda, I)$ is the likelihood function, i.e.\ the probability of the data being accurately described by a certain parameter choice. Assuming Gaussian noise, $\textbf{n}(t)$, we have
\begin{equation}
    \mathcal{L}( \textbf{d}| \blambda, I) \propto \exp\left[-\frac{1}{2} \langle \textbf{d}(t) - \textbf{h}(t, \blambda) \, | \, \textbf{d}(t) - \textbf{h}(t, \blambda) \rangle \right],
\end{equation}
where $\langle \cdot | \cdot \rangle$ denotes the noise-weighted inner product defined as 
\begin{equation}
\langle a | b \rangle = 2\, \int_{f_\mathrm{low}} ^{f_\mathrm{high}} \frac{\tilde{a}^*(f)\tilde{b}(f) + \tilde{a}(f)\tilde{b}^*(f)}{S_n(f)} \, \mathrm{d}f.
\end{equation}
Here $\tilde{a}(f)$, $\tilde{b}(f)$ represent the Fourier transform of $a(t)$, $b(t)$ and $f_\mathrm{low}$, $f_\mathrm{high}$ indicate the frequency cut-offs used in analysing the data which usually coincides with the bandwidth of the gravitational-wave detector's sensitivity. In practice, $P(\blambda | \textbf{d}, I)$ is sampled over the parameter space stochastically using Metropolis--Hastings Markov-Chain Monte Carlo~\cite{Metropolis:1953am,Hastings:1970aa} or nested sampling~\cite{Skilling2004a}. Considering the results that we use later, Refs.~\cite{GW170817,Johnson-McDaniel:2021yge,Narayan:2023vhm} used the LALInference code~\cite{Veitch:2014wba}, a part of LVK's LALSuite software library~\cite{lalsuite} while Refs.~\cite{GW190412,GW190521,GW190814,O3a_TGR,GWTC-3_paper,O3b_TGR} used both LALInference and Bilby~\cite{Ashton:2018jfp} to construct the posteriors, $P(\blambda | \textbf{d}, I )$.

\subsection{Inference using the GR model}

As mentioned above, we need a hypothesis or waveform model to describe our gravitational-wave signal. For want of a credible alternative, our dominant hypothesis is GR, and any investigation into beyond-GR effects starts with setting a GR baseline. 

There are three unique phases of a BBH merger: an initial \emph{inspiral} where the two black holes slowly coalesce through the back-reaction of gravitational-wave emission, the actual process of \emph{merger} whose onset is indicated by the formation of the apparent horizon, and the post-merger \emph{ringdown} where the newly formed remnant object settles down to a stable state through the emission of (a superposition of) exponentially damped sinusoidal gravitational waves, the quasi-normal-mode spectrum.

In GR, the entire gravitational-wave signal from a quasi-circular BBH merger is completely described by the masses and spins of the initial black holes, $(m_1, m_2, \boldsymbol{\chi}_1, \boldsymbol{\chi}_2)$, its position in the sky $(\alpha, \delta)$ and orientation $(\iota, \psi)$, luminosity distance $\dl$ and two reference quantities for phase and time $(\phi_c, t_c)$. Together, they form the 15 dimensional parameter space $\blambda_{\text{GR}} = \{m_1, m_2, \boldsymbol{\chi}_1, \boldsymbol{\chi}_2, \alpha, \delta, \dl, \iota, \psi,\phi_c, t_c\}$. Setting the GR baseline therefore involves performing parameter inference of $\blambda_{\text{GR}}$ with a GR waveform model. As mentioned earlier, these analyses can use a variety of different waveform models with varying assumptions. In this paper we use results from analyses with IMRPhenomPv2~\cite{Hannam:2013oca,Khan:2015jqa,Bohe:PPv2} for the GW150914-like and GW170608-like simulated observations and from analyses with IMRPhenomXPHM~\cite{Pratten:2020ceb} for the quasicircular and eccentric simulated observations using numerical relativity waveforms. Both of these models describe the waveforms from quasicircular precessing binary black holes, with IMRPhenomXPHM including two-spin precession effects and higher-order modes. We also use results obtained with the slightly earlier precessing model with higher modes, IMRPhenomPv3HM~\cite{Khan:2019kot} for some real events, and with the precessing binary neutron star model IMRPhenomPv2\_NRTidal~\cite{Dietrich:2017aum,Dietrich:2018uni} for GW170817.

In our study, we are particularly interested in the mass $\mf$ and spin $\af$ of the remnant black hole. Analytical fits to numerical relativity simulations can translate our estimates of the initial masses and spins $(m_1, m_2, \boldsymbol{\chi}_1, \boldsymbol{\chi}_2)$ into the estimates of the final mass and spin $(\mf, \af)$. The different waveform models incorporate different fits to predict the final mass and spin. We instead use a single set of fits in this analysis (the same used by the LVK to infer the final mass and spin in, e.g.~\cite{GWTC-3_paper}) to avoid any biases due to different final mass and spin fits. Specifically, we use an average of fits to numerical relativity simulations~\cite{Hofmann:2016yih,Healy:2016lce,Jimenez-Forteza:2016oae} where the aligned-spin final spin fits are augmented with the contribution from in-plane spin components~\cite{spinfit-T1600168} for precessing waveform models. However, unlike the application in \cite{GWTC-3_paper}, we do not evolve the initial spins before applying the fits; this is the same procedure used in the computation of the final mass and spin for the standard IMRCT in the LVK test of GR analyses~\cite{O2_TGR, O3a_TGR, O3b_TGR}. Since these fits are just used to provide a physically motivated two-dimensional parameter space in which to compare the results, the small decrease in accuracy of the fits due to not evolving the spins is not a concern.

While there are alternative theories of gravity, their credibility is contingent on their mathematical well-posedness, as well as their consistency with observations from gravitational waves and other astrophysical or cosmological observations. Hence, most strong-field tests of gravity are essentially tests of GR, where we measure and constrain possible deviations from GR predictions. Such tests are called null tests, GR being the null hypothesis. Three of the most prominent null tests of GR using gravitational-wave observations of BBH mergers are the IMRCT, parameterised tests of the gravitational waveform and the test of the modified dispersion relation, as discussed below.

\subsection{IMR Consistency Test}
\label{ssec:imrct}

The three phases of a BBH merger, as described above, are conventionally pictured in the time domain (as in figure~2 of~\cite{LIGOScientific:2016lio}). However, under the stationary phase approximation, the early (late) portions of the time-domain signal corresponds, to a high degree of accuracy, to the low- (high-) frequency portion of the Fourier-domain signal, as illustrated in \cite{Ghosh:2017gfp}. We thus refer to these portions of the Fourier-domain signal as the inspiral and the merger-ringdown portions and refer to the dividing frequency as $f_{\text{cut}}$. [The merger-ringdown portion is referred to as the post-inspiral portion in the LV(K) testing GR catalogue papers~\cite{O2_TGR, O3a_TGR, O3b_TGR}.]

The IMRCT infers the mass and spin of the remnant black hole, $(\mf, \af)$ from the low- and high-frequency portions of the signal, and compares these estimates with each other. If the entire gravitational-wave signal is accurately described by our GR hypothesis (and specifically by the family of waveforms used to carry out the test), then the low- and high-frequency inferences should correspond to the same underlying signal, or in terms of the remnant black hole, consistent measurements of $(\mf, \af)$. In order to quantify possible deviations between the two measurements, we define fractional deviations in the measurement of these quantities as:
\begin{subequations}
\begin{align}
    \dmfbymf &= 2\frac{\mf^{\text{I}} - \mf^{\text{MR}}}{\mf^{\text{I}} + \mf^{\text{MR}}}, \\
    \dafbyaf &= 2\frac{\af^{\text{I}} - \af^{\text{MR}}}{\af^{\text{I}} + \af^{\text{MR}}},
\end{align}
\end{subequations}
where the ``I'' and ``MR'' superscripts refer to the inspiral and merger-ringdown portions of the signal.
If GR is correct, the posteriors on $(\Delta M_{\text{f}}/\bar{M}_{\text{f}}, \Delta \chi_{\text{f}}/\bar{\chi}_{\text{f}})$ will be consistent with $(0,0)$.

The IMRCT takes $f_{\text{cut}}$ to be the $|m|=2$ gravitational-wave frequency of the innermost stable circular orbit (ISCO) of the remnant Kerr black hole~\cite{Bardeen:1972fi} computed from GR analysis of the full signal. The LVK analyses~\cite{O2_TGR, O3a_TGR, O3b_TGR} first compute the medians of the individual masses and spherical coordinate components of the spins from the GR analysis of the full signal and then use these values to compute a value for the final mass and spin and thence the ISCO frequency. The analysis in \cite{Narayan:2023vhm} computes the median of the ISCO frequency of the remnant Kerr black hole inferred from the GR analysis, but also checks that for the cases they consider the LVK method would give a negligible difference in $f_{\text{cut}}$.

The IMRCT can be performed using any GR waveform model. The analyses considered here use the IMRPhenomPv2 and IMRPhenomXPHM models.

\subsection{Parameterised tests}
The parameterized tests are designed to identify a GR violation by allowing deviations in the frequency-domain phasing coefficients of GR waveform models. The frequency-domain phase of the waveform in the early inspiral can be expressed analytically in a closed form using the stationary phase approximation as
\begin{equation}
\Phi(f)=\frac{3}{128\eta v^5}\sum_{k=0}^7 (\varphi_k v^k + 3\varphi_{kl}v^k \ln v),
\end{equation}
where $\varphi_k$ and $\varphi_{kl}$ are coefficients in the post-Newtonian (PN) expansion, and this expression extends to $3.5$PN, the highest order known completely at the time the tests in question and the GR waveform models they are based on were finalized. All of the $\varphi_k$ in the sum are nonzero, in general, except for $\varphi_1$, while the $\varphi_{kl}$ coefficients of the logarithmic terms are only nonzero for $k\in\{5, 6\}$ up to the PN order considered here. In the above expression, $\eta= m_1 m_2/(m_1+m_2)^2$ denotes the symmetric mass ratio and $v = (\pi M f)^{1/3}$, with $M= m_1 + m_2$ being the binary's total redshifted mass and $f$ being the gravitational-wave frequency. The PN expressions are not sufficient to describe the late-inspiral and merger-ringdown stages of the signal, which are instead described in different ways in different GR waveform models. For the IMRPhenomD model~\cite{Khan:2015jqa} relevant here, they are described by analytical expressions using the phenomenological coefficients $\beta_k$ and $\alpha_k$ that are fit to numerical relativity simulations. If $p_k$ collectively denotes all of the inspiral and merger-ringdown phasing coefficients $\varphi_k$, $\varphi_{kl}$, $\beta_k$ and $\alpha_k$, then deviations from GR at all stages can be introduced by inserting the dimensionless fractional deformation parameters $\delta\hat{p}_k$ into the phase such that $p_k \to (1 + \delta\hat{p}_k)p_k$. If the observed gravitational-wave signal is consistent with GR and more specifically described by the GR waveform family used to perform the analysis, all $\delta\hat{p}_k$ should to be consistent with zero.

As mentioned earlier there are two varieties of the parametrized test. The first one is the Test Infrastructure for General Relativity (TIGER) approach~\cite{Agathos:2013upa,Meidam:2017dgf} in which the deviations are introduced only in the non-spinning part of the PN coefficients $\varphi_k$, $\varphi_{kl}$, as well as in the phenomenological $\beta_k$ and $\alpha_k$ coefficients. The restriction to modifying the non-spinning part of the PN coefficients prevents situations where the spin terms cancel the nonspinning terms in the PN coefficient, making the parameterisation singular. The LVK analyses so far have used a TIGER version that introduces the deviations in the phase of aligned-spin dominant mode waveform model IMRPhenomD~\cite{Khan:2015jqa} and its tidal extension IMRPhenomD\_NRTidal~\cite{Dietrich:2017aum,Dietrich:2018uni}. This modified IMRPhenomD (IMRPhenomD\_NRTidal) phase is then twisted up to produce a modified precessing IMRPhenomPv2 (IMRPhenomPv2\_NRTidal) waveform model using GR spin precession equations.
The second variety is the Flexible-Theory-Independent (FTI) method~\cite{Mehta:2022pcn} that considers deviations in only the inspiral PN coefficients (including the spin contributions) at low frequencies and tapers those deviations to zero at a specific cutoff frequency, unlike TIGER that allows inspiral deviations to affect the post-inspiral regime through the $C^1$ matching used to construct the IMRPhenomD model. The FTI test can be implemented based on any aligned-spin GR waveform model, but the version used in the LVK analyses~\cite{GW170817_TGR,O2_TGR, O3a_TGR, O3b_TGR} is based on the effective-one-body reduced-order model SEOBNRv4\_ROM~\cite{Bohe:2016gbl} and its higher-order mode version~\cite{Cotesta:2018fcv,Cotesta:2020qhw} or tidal version SEOBNRv4\_ROM\_NRTidal~\cite{Dietrich:2017aum,Dietrich:2018uni}. The FTI results are reweighted to the TIGER convention, as in the LVK analyses~\cite{GW170817_TGR,O2_TGR, O3a_TGR, O3b_TGR}, so that the results of the two analyses are directly comparable.

\subsection{Test of the modified dispersion relation}
The modified dispersion relation (MDR) test of gravitational waves constrains dispersive propagation of gravitational waves, specifically considering a phenomenological modification of the standard dispersion relation. Gravitational waves in GR propagate non-dispersively as described by the standard dispersion relation $E^2 = p^2$,  where $E$ and $p$ denote the energy and momentum of the wave. The dispersion relation
\begin{equation}
E^2 = p^2 + A_\alpha p^\alpha
\end{equation}
introduced in \cite{Mirshekari:2011yq} includes the leading predictions of a number of non-GR theories, as discussed in that paper and~\cite{O2_TGR}. Here $A_\alpha$ and $\alpha$ are phenomenological parameters, where $A_\alpha$ determines how much the phase of the gravitational-wave signal deviates from the predictions of GR and $\alpha$ determines the frequency dependence of the frequency-domain dephasing, which is $\propto f^{\alpha - 1}$ (with a logarithmic expression for $\alpha = 1$ in some implementations); see the explicit expressions in~\cite{Mirshekari:2011yq,O2_TGR}. Here one assumes that the deviation from GR only comes from the propagation of the waves and not from its generation, i.e.\ the signal close to the source is still consistent with GR to a very good approximation, which is likely to be the case for at least some of the alternative theories that predict modified propagation, notably massive graviton theories, as discussed in~\cite{O2_TGR}.

In GR, $A_\alpha=0$ for all $\alpha$, but in a beyond GR theory they can take nonzero values. For example, in a massive graviton theory $\alpha=0$ and $A_0>0$. The LVK analyses~\cite{GW170817_TGR, O2_TGR, O3a_TGR, O3b_TGR} consider $\alpha\in\{0, 0.5, 1, 1.5, 2.5, 3, 3.5, 4\}$ ($\alpha=2$ is omitted since it corresponds to no dispersion of the wave), and we consider these values or a subset here: The analyses of the GR and non-GR waveforms from~\cite{Johnson-McDaniel:2021yge} just considers $\alpha = 0$ and the analysis of numerical relativity waveforms in~\cite{Narayan:2023vhm} omits $\alpha = 1$ \footnote{The analyses in~\cite{Narayan:2023vhm} omit $\alpha=1$ since the implementation of the test they use gives the logarithmic frequency dependence corresponding to using the particle velocity of the wave, while the group velocity that is physically preferable gives a constant dephasing (which still leads to a non-GR signal in the presence of higher modes). We still consider the $\alpha = 1$ cases in the application of the meta IMRCT to real events, since it is still a fine phenomenological test of GR in this case, even though it loses the motivation of coming from the dispersion relation.}. As described in~\cite{O2_TGR}, the analysis is carried out by sampling over an effective wavelength parameter,
considering positive and negative values of $A_\alpha$ separately. The results from two signs of $A_\alpha$ are then combined and reweighted to a flat prior in $A_\alpha$ to obtain the final posterior on $A_\alpha$ for different values of $\alpha$.

\section{Meta IMR consistency test}
\label{sec:meta_imrct}

The meta IMRCT is a straightforward extension of the standard IMRCT described in section~\ref{ssec:imrct}. Instead of using the final mass and spin inferred from the low- and high-frequency portions of the analysis of a signal, one instead uses the final mass and spin inferred using two different tests of GR, or a test of GR and the standard GR analysis of the signal (henceforth referred to as GR parameter estimation [PE]). Specifically, one has
\begin{subequations}
\begin{align}
    \dmfbymf &= 2\frac{\mf^{T} - \mf^{T'}}{\mf^{T} + \mf^{T'}}, \\
    \dafbyaf &= 2\frac{\af^{T} - \af^{T'}}{\af^{T} + \af^{T'}},
\end{align}
\end{subequations}
where $T$ and $T'$ are two separate analyses.

In this application, we break down each test of GR into its individual runs and apply the meta IMRCT to these. Specifically, we consider the inspiral and merger-ringdown analyses used in the standard IMRCT separately. Thus, the standard IMRCT is a special case of the meta IMRCT, though we exclude the standard IMRCT from our meta IMRCT results and just consider pairs of inspiral or merger-ringdown analyses with other tests or GR PE. For the TIGER, FTI, and modified dispersion analyses, we consider pairs involving all the different testing parameters separately, both with other testing parameters from the same test as well as all other tests.

While the standard IMRCT is restricted to binaries in a certain mass range where there is sufficient signal-to-noise (SNR) in both portions of the signal (see, e.g.\ \cite{O3b_TGR}), due to its analysis of the low- and high-frequency portions of the signal independently, the meta IMRCT has no such restrictions. The meta IMRCT can even be applied to binary neutron star or neutron star--black hole binary signals: One can even apply the same BBH fits in this case when applying the test, since the fits just provide a convenient way of mapping the initial mass and spin parameter space to a two-dimensional space. In fact, one does not have to apply the test to the final mass and spin, but could choose other GR parameters to compare, which we briefly explore in section~\ref{sssec:luminosity-distance}. However, we leave a full exploration of alternative parameters to future work. The meta IMRCT is also computationally quite efficient, once one has the results from the individual tests, since it does not require any further stochastic sampling or evaluation of waveforms.

As mentioned previously, one expects the meta IMRCT to be sensitive to deviations from GR (or more generally, from the waveform model used to perform the analysis), given the results in~\cite{Johnson-McDaniel:2021yge}, which show that the different tests (or GR PE) have different biases in the final mass and spin when applied to non-GR signals (see specifically figures 10--12 in that paper). This is as one would expect: When one performs a test of GR or GR PE on a signal that is not described by the underlying GR waveform model used to perform the analysis, one will obtain different posteriors on the individual masses and spins depending on the deviations from GR allowed in the analysis. These differences in the individual masses and spins will, in general, translate to different final masses and spins.

As is done for the standard IMRCT, we reweight the posterior on the deviation parameters $(\Delta M_f/\bar{M}_f, \Delta\chi_f/\bar{\chi}_f)$ to a flat prior by dividing by the distribution of the deviation parameters induced by the prior distributions for the individual analyses. To have a single number to measure the consistency of the meta IMRCT result with GR, we compute the GR quantile $Q_\text{GR}$, which is the fraction of the reweighted posterior probability contained within the isodensity contour that intersects the GR value of $(0,0)$, or
\<
Q_\text{GR} = \int_{P(\delta_M,\delta_\chi) > P(0,0)}P(\delta_M,\delta_\chi)\text{d}\delta_M\text{d}\delta_\chi,
\?
where $P$ is the reweighted posterior probability density and $\delta_M := \Delta M_f/\bar{M}_f$, $\delta_\chi := \Delta\chi_f/\bar{\chi}_f$.
Thus, small GR quantiles correspond to agreement with GR, while GR quantiles close to $100\%$ correspond to significant GR deviations. We also sometimes quote the Gaussian $\sigma$ equivalent to the GR quantile, i.e.\ the number of $\sigma$s away from the mean at which the probability contained within a symmetric interval with edges that distance from the mean is the same as the GR quantile, or symbolically, $x$, where
\<
Q_\text{GR} = \int_{-x}^x\mathcal{N}(\bar{x})\text{d}\bar{x},
\?
with $\mathcal{N}$ the standard normal distribution.

\subsection{Effect of priors and null distribution in Gaussian noise}
\begin{figure}[htb]
\centering
\includegraphics[width=.45\textwidth]{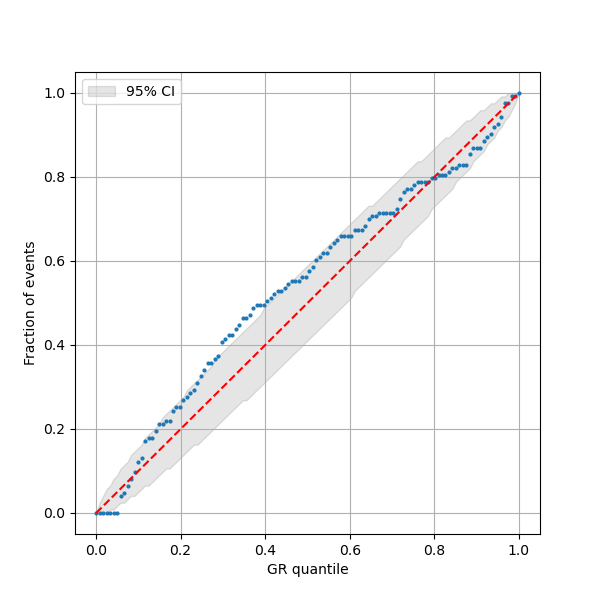}
\includegraphics[width=.45\textwidth]{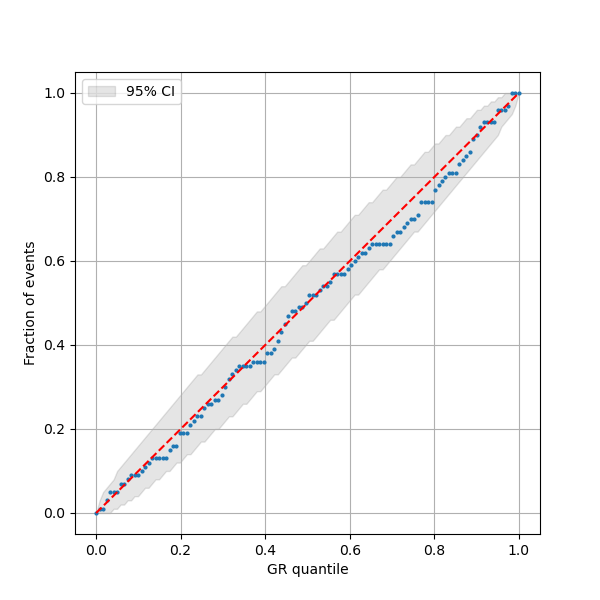}
\caption{\label{fig:pp}P-p plots for the standard IMRCT analysis of simulated observations in Gaussian noise (left) and the meta IMRCT applied to mock data with different prior ranges for the two analyses (right). The red straight line in both the plots represents theoretical expectation. In both cases, we show the $95\%$ bound on the expected statistical variation due to the finite number of simulated observations considered. The red dashed line represents the theoretical expectation.}
\end{figure}

One subtle point in interpreting the results of the meta IMRCT is its dependence on two prior probability distributions. (This also affects the standard IMRCT, but there it is customary to choose the same prior distributions for both the inspiral and merger-ringdown analyses.) In particular, if the priors on the masses and/or spins for the individual analyses are different, this can lead to a prior on the IMRCT deviation parameters that peaks away from the GR value of zero for one or both parameters. For instance, the prior on the deviation parameters peaks well away from the origin for several pairs of GW190412 and GW190814 analyses, due to different choices for the mass prior ranges, as discussed in the next section.

However, the reweighting to a flat prior in the deviation parameters we perform should remove these prior effects. We check this explicitly in two ways, deferring to future work a direct test of the meta IMRCT applied to a variety of pairs of tests applied to simulated observations in Gaussian noise, due to the considerable computational expense of such a study. First, we consider the standard IMRCT applied to the set of simulated observations in Gaussian noise considered in~\cite{Ghosh:2017gfp} and show that the reweighting produces an approximately flat distribution of GR quantiles, as is illustrated by the p-p plot in the left panel of figure~\ref{fig:pp}. There is more deviation from the diagonal in the p-p plot than would be expected due to the number of simulated observations, so we also assess the extent to which these differences from an exactly uniform distribution of GR quantiles would affect our application of the meta IMRCT to real events (the only ones for which noise is relevant) in section~\ref{ssec:real_events}. The specific simulated observations we consider from~\cite{Ghosh:2017gfp} are a set of $219$ aligned-spin signals. For the purposes of this check, we do not make the restriction to total masses $< 150M_\odot$ made in that paper. However, we do remove all the simulated observations where there is nonnegligible railing against the prior boundaries in the parameter estimation results, leaving us with $123$ observations, and reweight to the flat prior in the deviation parameters. Thus, our results are not the same as those in the left-hand panel of figure~4 in~\cite{Ghosh:2017gfp}.

The simulated observations in Gaussian noise from~\cite{Ghosh:2017gfp} were analyzed with the same priors for both the inspiral and merger-ringdown. We thus also explicitly check a situation where the priors in the two analyses are not the same, as one will encounter in applications of the meta IMRCT. Since performing many simulated observations is computationally expensive, we instead use mock data generated following~\cite{Farah:2023vsc}. Applying the meta IMRCT to the mock data produces the p-p plot in the right panel of figure~\ref{fig:pp}, which is nicely in agreement with the expectation for a uniform distribution of GR quantiles, within the uncertainties due to the $100$ cases we used to make the plot. Specifically, we generate a population of binaries with redshifted chirp masses and mass ratios distributed uniformly in $[26,29]M_\odot$ and $[0.3, 1]$, respectively, with uniform distributions of spin magnitudes and spin angles (uniform on the $2$-sphere). (While we use the convention that the mass ratio is $\geq 1$ in the rest of the paper, in this subsection we use the $\leq 1$ convention, since this is standard in GW parameter estimation.) The mock data are then generated by taking uncorrelated Gaussian distributions in each quantity, where the log of the chirp mass $\ln\mathcal{M}$ and the symmetric mass ratio $\eta$ are used for those quantities, and the data are then reweighted to a flat prior in the individual masses, as in~\cite{Farah:2023vsc}. The medians of the Gaussians are shifted so that the distribution of the quantiles of the true values is flat. The standard deviations of the Gaussians are all scaled by $1/\text{SNR}$, where the SNR is drawn from the standard $1/\text{SNR}^4$ distribution with a minimum value of $10$. The values of the standard deviations at $\text{SNR} = 10$ are $\{0.0067, 0.0067, 0.1, 0.1, 0.6, 0.6, 1\}$ and $\{0.01, 0.01, 0.2, 0.3, 0.8, 1.1, 0.8\}$ for the two cases ($T$ and $T'$ in the meta IMRCT notation above) for the parameters $\{\ln\mathcal{M}, \eta, \chi_1, \chi_2, \cos\theta_1, \cos\theta_2, \phi_{12}\}$. Here $\theta_{1,2}$ are the (polar) angles between the spin vectors and the binary's (Newtonian) orbital angular momentum (i.e.\ the normal to the orbital plane), while $\phi_{12}$ is the difference between the azimuthal angles of the spins---the remaining spin angle is not used in the meta IMRCT, since it does not enter into the fits for the final mass and spin we use. We consider chirp mass and mass ratio prior ranges of $\{[24,31]M_\odot, [0.15, 1]\}$ and $\{[22,32]M_\odot, [0.1, 1]\}$ for the two cases, along with the standard restriction of $[1,100]M_\odot$ for the individual masses used in the LALInference parameter estimation code~\cite{lalsuite} and the standard uniform spin priors. The specific values we use for the range of masses, uncertainties, and prior ranges were chosen quasi-randomly while still avoiding nonnegigible railing at the prior boundaries. We checked that the p-p plots for the individual parameters are also in agreement with the expectation within the uncertainties due to the number of simulated observations.

\subsection{Combined $p$-value}

The meta IMRCT described above produces $N(N-1)/2$ results when applied to $N$ tests of GR. In order to have a single number that describes the consistency with GR for each application of the meta IMRCT, we use the modified Bonferroni procedure due to Simes~\cite{Simes} to compute a combined $p$-value. Specifically, in this procedure, the combined $p$-value is given by $p = N\min_j P_{(j)}/j$, where $P_{(1)} \leq P_{(2)} \leq \cdots \leq P_{(N)}$ are the ordered $p$-values from the application of the meta IMRCT to the $N$ individual pairs. The $p$-value for a single pair is given by $1 - Q_\text{GR}$. We use the Simes method rather than the classical Bonferroni procedure of introducing a trial factor (also reviewed by Simes~\cite{Simes}) since the latter is overly conservative (i.e.\ loses significant statistical power) in cases like the meta IMRCT, where the individual tests are highly (positively) correlated. However, the Simes method requires additional assumptions about the correlation of the tests, and while it is conservative in many cases with positive correlation, there can be a weakening of up to at least a factor of $2$ in some positively correlated cases~\cite{Gou03092023}. As shown by Hommel~\cite{Hommel1983tests}, one can obtain a completely conservative result by including the factor $C_N = \sum_{j=1}^N(1/j)$, i.e.\ quoting $\min(C_N p,1)$. However, $C_N$ grows roughly as $\ln N$, so this gives a significantly weaker test than the pure Simes test. For the current work, we have not tried to establish whether any of the known (but not particularly easy to check) hypotheses about correlation that guarantee the strict validity of the Simes test (discussed in, e.g.~\cite{Gou03092023}) are satisfied by the meta IMRCT. Thus, we also quote the Hommel upper bound, for comparison, since it is still much more constraining than the classical Bonferroni procedure for our results, as we illustrate in section~\ref{sec:results}.

\section{Simulated observations}
\label{sec:sim_obs}

We consider two sets of simulated observations: GR and non-GR simulated observations like GW150914 and GW170608 from~\cite{Johnson-McDaniel:2021yge} and quasicircular and eccentric simulated BBH observations (using numerical relativity simulations) from~\cite{Narayan:2023vhm}. We use the results of parameter estimation with GR waveforms (GR PE) and applications of the tests of GR described in section~\ref{sec:tests} obtained in these previous papers. All of these cases do not include noise in the simulated observations (i.e.\ consider the zero realization of the noise) to avoid biases from a specific nonzero noise realization.

\subsection{GW150914-like and GW170608-like GR and non-GR cases}

We first detail the simulated observations from~\cite{Johnson-McDaniel:2021yge}, which are designed to emulate the GW150914~\cite{GW150914} and GW170608~\cite{GW170608} events, as fiducial high- and low-mass BBHs. These use both GR and non-GR waveforms, considering both large and small deviations in the GW150914-like case, and only small deviations in the GW170608-like case. The GR waveforms considered are the Institut des Hautes {\'E}tudes Scientifiques (IHES) effective-one-body (EOB) model~\cite{Damour:2012ky} and the IMRPhenomD phenomenological model~\cite{Khan:2015jqa}, and just include the dominant $\ell = |m| = 2$ modes of the waveform. The IHES EOB model is used to obtain a non-GR waveform by modifying its energy and angular momentum flux by multiplying the modes of the waveform that first enter these at $2$PN by a factor $\sqrt{a_2}$. Thus, the inspiral phase is also modified starting at $2$PN. Additionally, the final mass and spin are set self-consistently to satisfy energy and angular momentum balance, so the merger-ringdown portion of the signal also differs from that in GR for the same initial masses (and spins). The IMRPhenomD model is the basis for a waveform with dispersive propagation from a massive graviton~\cite{Will:1997bb} with graviton mass $m_g$, leading to a dephasing of the entire waveform in the frequency domain $\propto m_g^2/f$ (as in the MDR test with $\alpha = 0$, as discussed above). It is also the basis for waveforms with the TIGER and FTI modifications described in section~\ref{sec:tests}, here considering the modification to the $2$PN phasing coefficient.

The (redshifted) total masses of all these simulated observations are $72.2M_\odot$ and $19.9M_\odot$ for the GW150914-like and GW170608-like cases. The mass ratio is $1.22$ for the EOB waveforms and $1$ for the others (due to a typo in generating the waveforms). The IHES EOB model does not allow for spin, so the spins are set to zero in these simulated observations (and both GW150914 and GW170608 are consistent with being nonspinning). These simulated observations have SNRs of $\sim 54$ (GW150914-like) and $\sim 21$ (GW170608-like) in the O3-like LIGO-Virgo sensitivities~\cite{Aasi:2013wya} considered, except for the modified EOB GW150914-like case, where the SNR is as small as $40$ (with the larger GR deviation).

The GR PE is carried out using the IMRPhenomPv2 waveform model, which also forms the basis for all the tests applied, except for FTI, where we use SEOBNRv4\_ROM as in most applications to LIGO-Virgo data~\cite{O2_TGR,O3a_TGR,O3b_TGR}. The applications of the TIGER and FTI tests to all these cases just consider the $2$PN deviation parameter, as well as the $1$PN deviation parameter in a few cases, and the $\delta\alpha_2$ and $\delta\beta_2$ post-inspiral parameters for TIGER. The MDR test just considers the $\alpha = 0$ case that agrees with the massive graviton dephasing.

The results use here include the corrected GW170608-like MDR simulated observations: The published results used an GW170608-like MDR simulated observation with no massive graviton dephasing, due to a typo in the command used to create it. An erratum for~\cite{Johnson-McDaniel:2021yge} is in preparation.

\subsection{Eccentric cases}

We now detail the simulated observations from~\cite{Narayan:2023vhm}, which use numerical relativity simulations of nonspinning BBHs from the Simulating eXtreme Spacetimes collaboration~\cite{Boyle:2019kee}, both quasicircular and with eccentricities of $\sim 0.05$ and $\sim 0.1$, at $\sim 17$~Hz for the total mass of $80M_\odot$ used for these cases (see~\cite{Hinder:2017sxy} for information about the eccentric simulations, including how the eccentricity is estimated). For each eccentricity, there are three mass ratios, $q$: $1$, $2$, and $3$. These simulated observations are created to be face-on, so only the $m = 2$ modes contribute, and the $\ell$ modes considered are then matched to the test being carried out. Thus, they are just the dominant $\ell = m = 2$ modes for TIGER and FTI, where the analyses use the IMRPhenomPv2 and SEOBNRv4HM\_ROM waveform models, respectively, but include the $\ell = 3$, $m = 2$ modes for MDR, which uses the IMRPhenomXPHM waveform model (and IMRCT, which also uses IMRPhenomXPHM, but we do not use those results here, as described below). The SNRs are $\sim 120$, $\sim 105$ and $\sim 90$ for mass ratios of $1$, $2$ and $3$, respectively, with the forecast O4 sensitivities~\cite{Aasi:2013wya} considered. We do not consider the IMRCT results in these cases, due to the bias found in~\cite{Narayan:2023vhm} for even the quasicircular cases, which is attributable to the face-on inclination angle and the inclusion of higher modes. 

\section{Results}
\label{sec:results}

We now apply the meta IMRCT to the simulated observations discussed in the previous section, using the results of the tests given in section~\ref{sec:tests} along with GR PE. Here we consider the positive and negative $A_\alpha$ results from the MDR test separately. Additionally, as mentioned above, we exclude the $(\text{I}, \text{MR})$ pair that just reproduces the standard IMRCT.

\begin{table*}[h!]

\caption[Summary of meta IMRCT on GW150914-like simulated observations]{\label{tab:GW150914-like}Summary of meta IMRCT on GW150914-like simulated observations (simul.\ obs.). Here $>$ represents for larger GR deviation and $<$ represents for smaller GR deviation. We show the specific testing parameters or other PE results used. We apply the meta IMRCT to all these pairs of results. The GR quantile and Gaussian $\sigma$ columns give the median and 90\% interval for the distribution of these quantities for all the pairs of tests considered. The GR quantile is two dimensional and $0$ denotes agreement with GR. We also give the combined $p$-value is computed using both the Simes method ($p_\text{S}$) as well as the more conservative Hommel method ($p_\text{H}$). The quantity $N^{100\%}$ denotes the number of pairs where the GR quantile from the meta IMRCT and the maximum GR quantile from the individual tests are both 100\%. The quantities $N_\text{mIMRCT}^{\text{grtr}, 90\%}$ and $N_\text{mIMRCT}^\text{grtr}$ denote the number of pairs where the meta IMRCT gives a larger GR quantile than the individual tests do, where the first one also restricts to the pairs where the meta IMRCT GR quantile is $\geq 90\%$. The quantity $N_\text{tot}$ gives the total number of pairs.}

\scalebox{0.66}{
\begin{tabular}{*{25}{c}}

\hline\hline
\\
\multirow{2}{*}{Simul.\ Obs.\ } & \multirow{2}{*}{GR PE} & \multirow{2}{*}{IMRCT }& \multirow{2}{*}{TIGER} & \multirow{2}{*}{FTI} & \multirow{2}{*}{MDR} & \multirow{2}{*}{GR Quantile (\%)} & \multirow{2}{*}{Gaussian $\sigma$} & Combined $p$-value &\multirow{2}{*}{$N^{100\%}/N_\text{mIMRCT}^{\text{grtr}, 90\%}/N_\text{mIMRCT}^\text{grtr}/N_\text{tot}$ }\\
&   &  & & & &  &  &$p_\text{S}, p_\text{H}$&\\
\\
\hline

EOB, GR & \checkmark & I, MR & $\delta \varphi_2, \delta \varphi_4,\delta \alpha_2,\delta \beta_2$&$\delta \varphi_2, \delta \varphi_4$ & $\tilde{A}_0$ & $2.8^{+63.1}_{-1.8}$ & $0.0^{+0.9}_{-0.0}$ & $0.991, 1$ & $0/0/12/54$\\[1pt]
\\
Phenom, GR &  \checkmark & I, MR & $\delta \varphi_2, \delta \varphi_4,\delta \alpha_2,\delta \beta_2$&$\delta \varphi_2, \delta \varphi_4$ & $\tilde{A}_0$ & $3.7^{+57.8}_{-3.1}$ & $0.0^{+0.8}_{-0.0}$ & $0.995, 1$ & $0/0/7/54$\\[2.5pt]
\\
modified EOB, $>$ &  \checkmark & I, MR & $\delta \varphi_2, \delta \varphi_4,\delta \alpha_2,\delta \beta_2$&$\delta \varphi_2, \delta \varphi_4$& $\tilde{A}_0$  &$99.6^{+0.4}_{-89.0}$ & $2.9^{+0.1}_{-2.8}$ &$6.23\times 10^{-3}, 2.85\times 10^{-2}$ &$25/2/2/54$\\[1pt]
\\
modified EOB, $<$& \checkmark &  I, MR & $ \delta \varphi_4,\delta \alpha_2,\delta \beta_2$&$\delta \varphi_4$ & $\tilde{A}_0$ &$12.4^{+58.8}_{-10.6}$ &$0.2^{+0.9}_{-0.1}$ & $0.992, 1$ &$0/0/5/35$\\[2.5pt]
\\
MDR, $>$& \checkmark &  I, MR & $\delta \varphi_2,   \delta \varphi_4,\delta \alpha_2,\delta \beta_2$&$\delta \varphi_2, \delta \varphi_4$& $\tilde{A}_0$ &$71.5^{+27.0}_{-68.7}$ & $1.1^{+1.4}_{-1.0}$ &$0.185, 0.848$ &$0/1/1/54$\\[1pt]
\\
MDR, $<$& \checkmark &  I, MR & $ \delta \varphi_4,\delta \alpha_2,\delta \beta_2$&$\delta \varphi_4$ & $\tilde{A}_0$ &$5.1^{+66.9}_{-3.8}$ & $0.1^{+1.0}_{-0.0}$ &$0.995, 1$ &$0/0/3/35$\\[2.5pt]
\\
TIGER, $>$& \checkmark &  I, MR & $\delta \varphi_4,\delta \alpha_2,\delta \beta_2$&$ \delta \varphi_4$& $\tilde{A}_0$  &$99.6^{+0.3}_{-98.4}$ & $3.0^{+0.1}_{-3.0}$ &$6.18\times 10^{-3}, 2.56\times 10^{-2}$ &$17/0/0/35$\\[1pt]
\\
TIGER, $<$& \checkmark &  I, MR & $\delta \varphi_4,\delta \alpha_2,\delta \beta_2$&$ \delta \varphi_4$& $\tilde{A}_0$ &$85.9^{+14.1}_{-84.4}$ & $1.5^{+1.5}_{-1.4}$ &$2.36\times 10^{-2}, 9.80\times 10^{-2}$ &$4/0/0/35$\\[2.5pt]
\\
FTI, $>$& \checkmark &  I, MR & $\delta \varphi_4,\delta \alpha_2,\delta \beta_2$&$ \delta \varphi_4$& $\tilde{A}_0$  & $100^{+0.0}_{-42.5}$ & $3.0^{+0.0}_{-2.1}$ &$3.50\times 10^{-3}, 1.45\times 10^{-2}$ &$30/0/0/35$\\[1pt]
\\
FTI, $<$ & \checkmark &  I, MR & $\delta \varphi_4,\delta \alpha_2,\delta \beta_2$&$ \delta \varphi_4$& $\tilde{A}_0$  &$90.5^{+9.4}_{-89.0}$ & $1.7^{+1.3}_{-1.6}$ &$1.05\times 10^{-2}, 4.35\times 10^{-2}$ &$10/1/1/35$\\[1pt]

\hline\hline
\end{tabular}
}
\end{table*}

\begin{figure*}[htb]
\centering
\subfloat{\includegraphics[width=.5\textwidth]{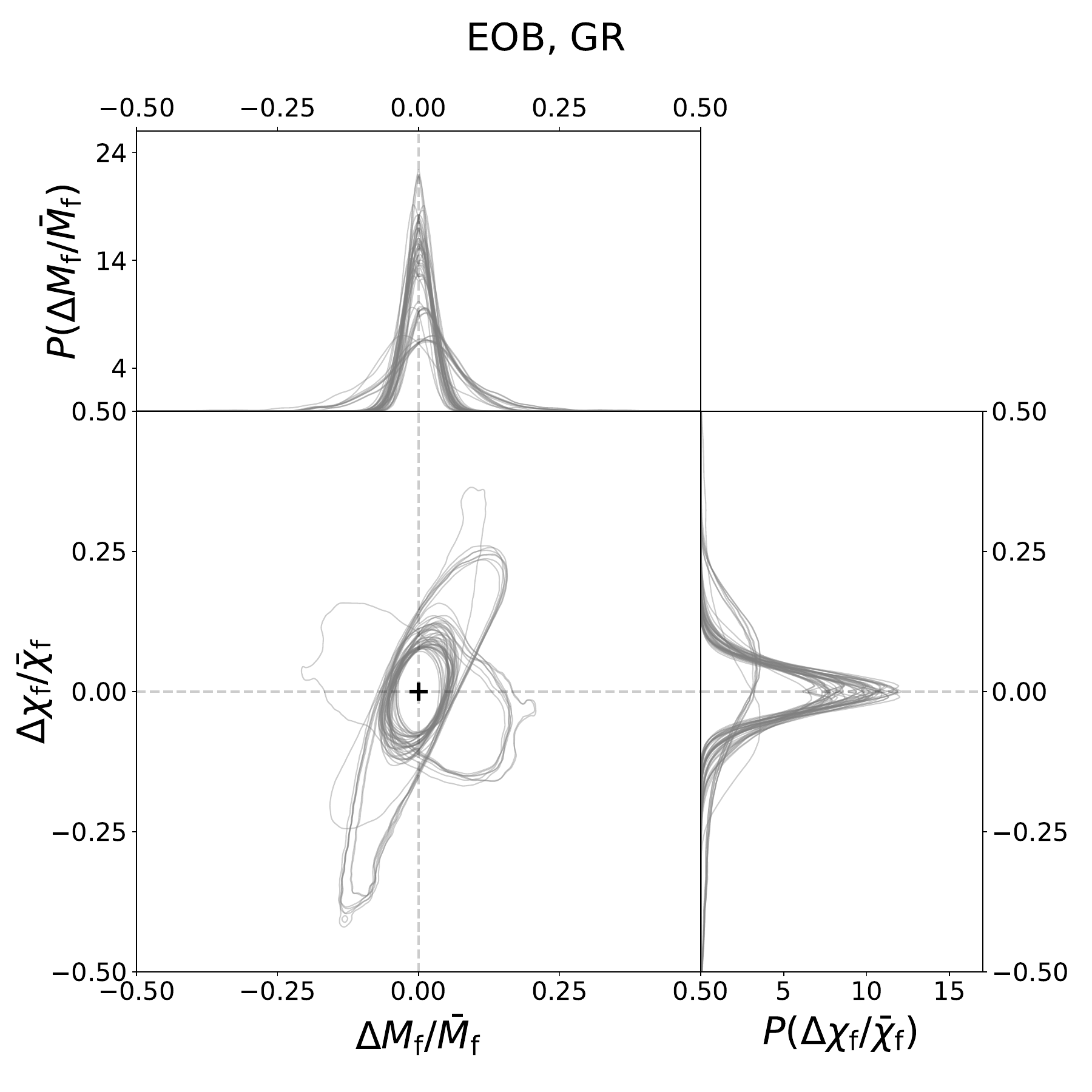}}
\qquad
\subfloat{\includegraphics[width=.5\textwidth]{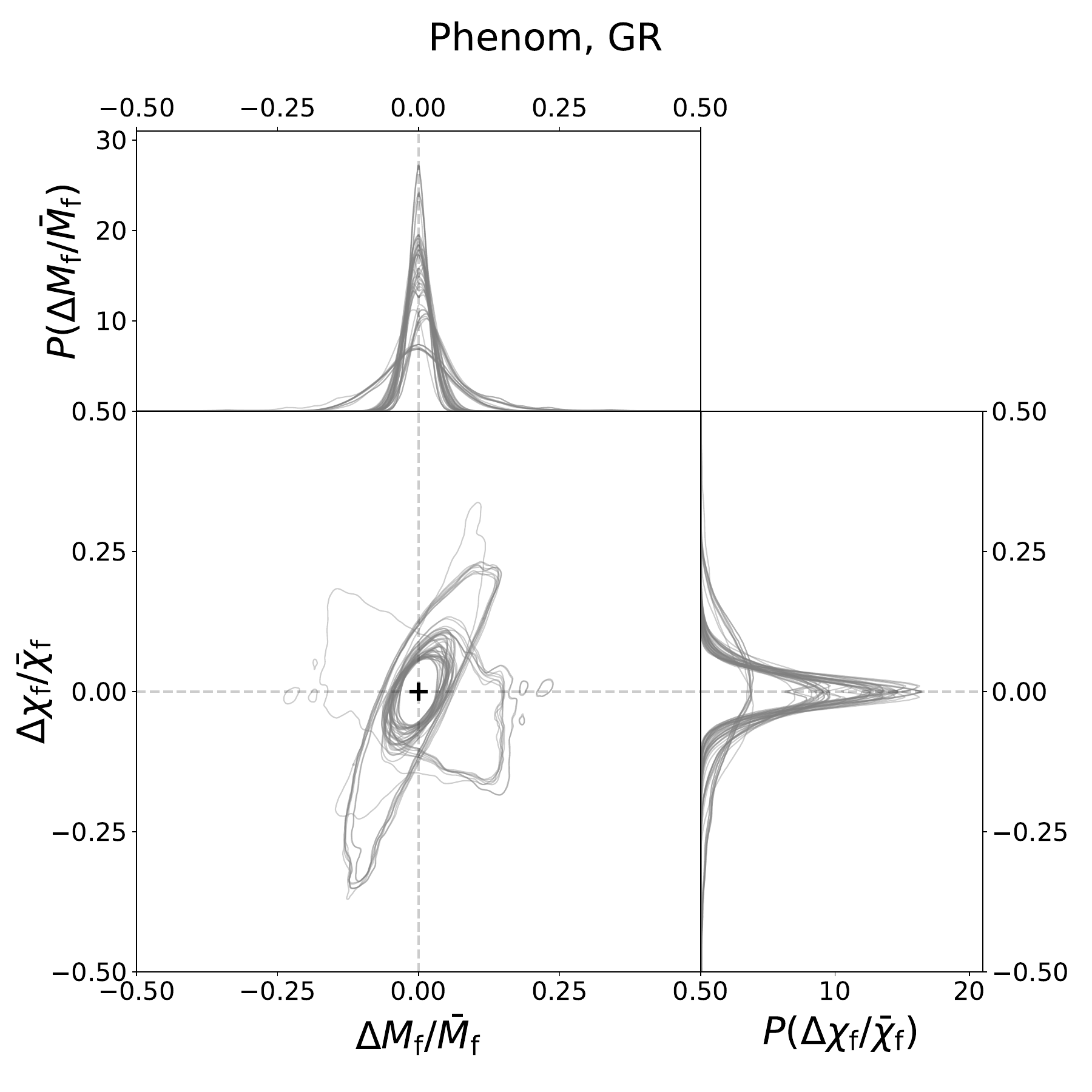}}\\

\subfloat{\includegraphics[width=.5\textwidth]{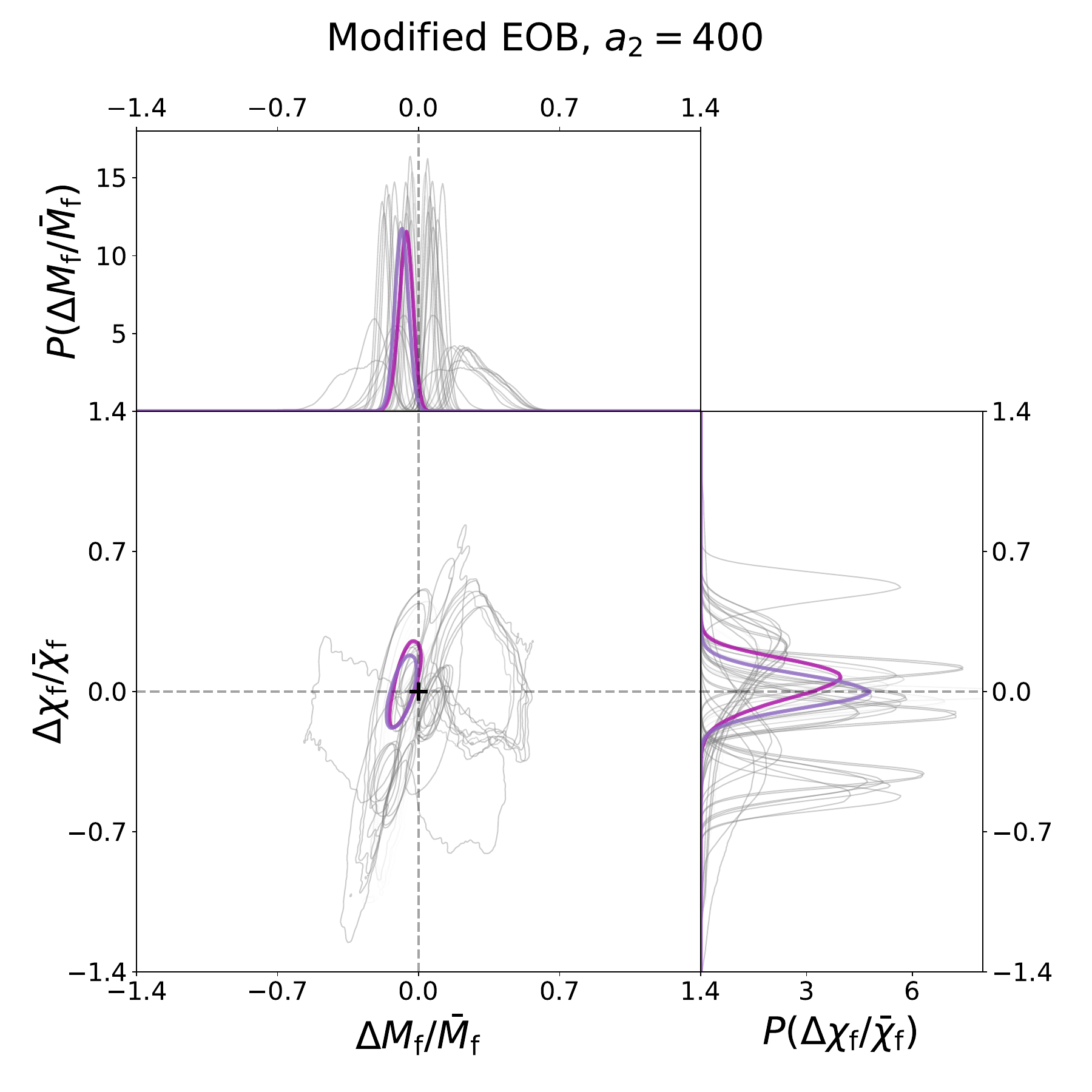}}
\qquad
\subfloat{\includegraphics[width=.5\textwidth]{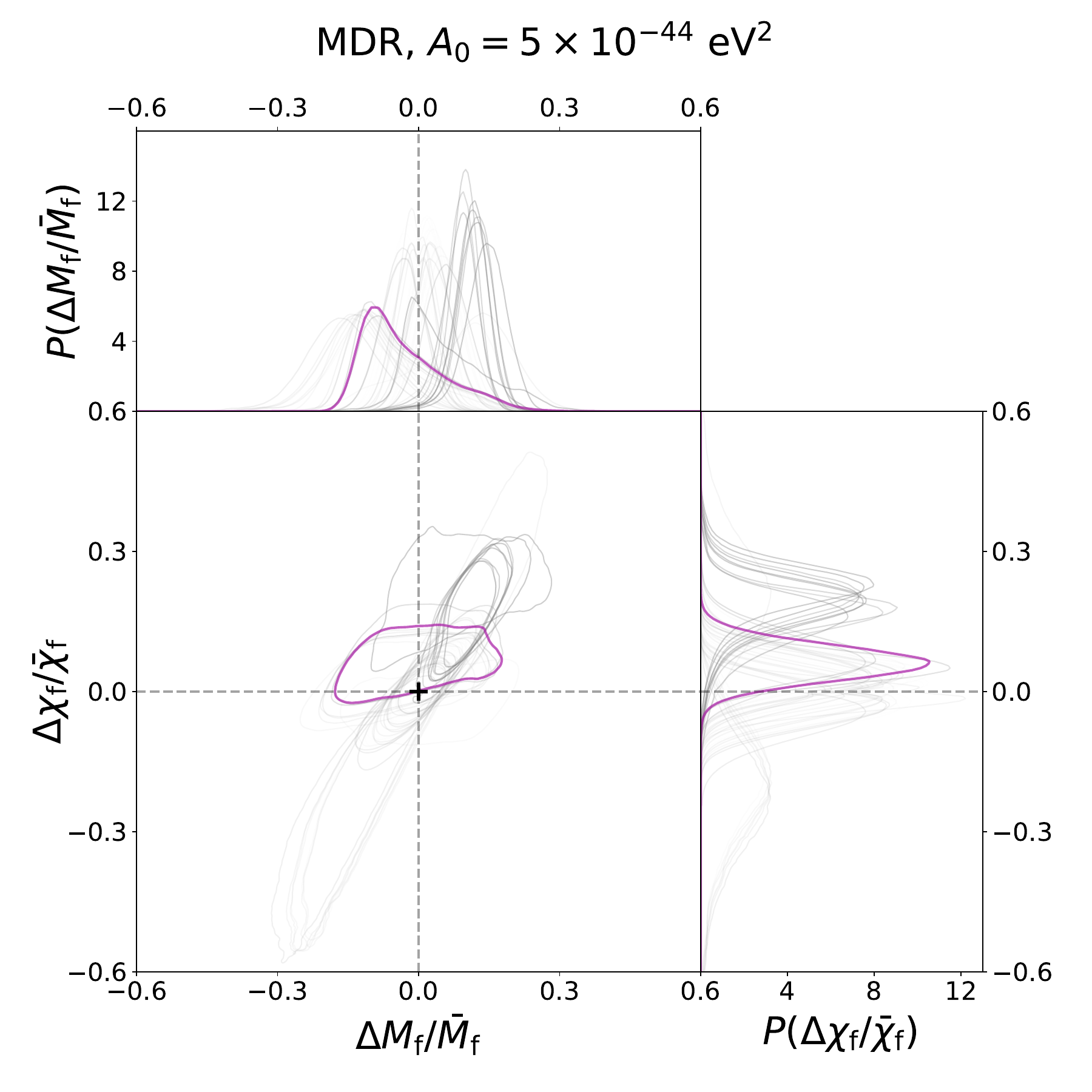}}\\

\subfloat{\includegraphics[width=.5\textwidth]{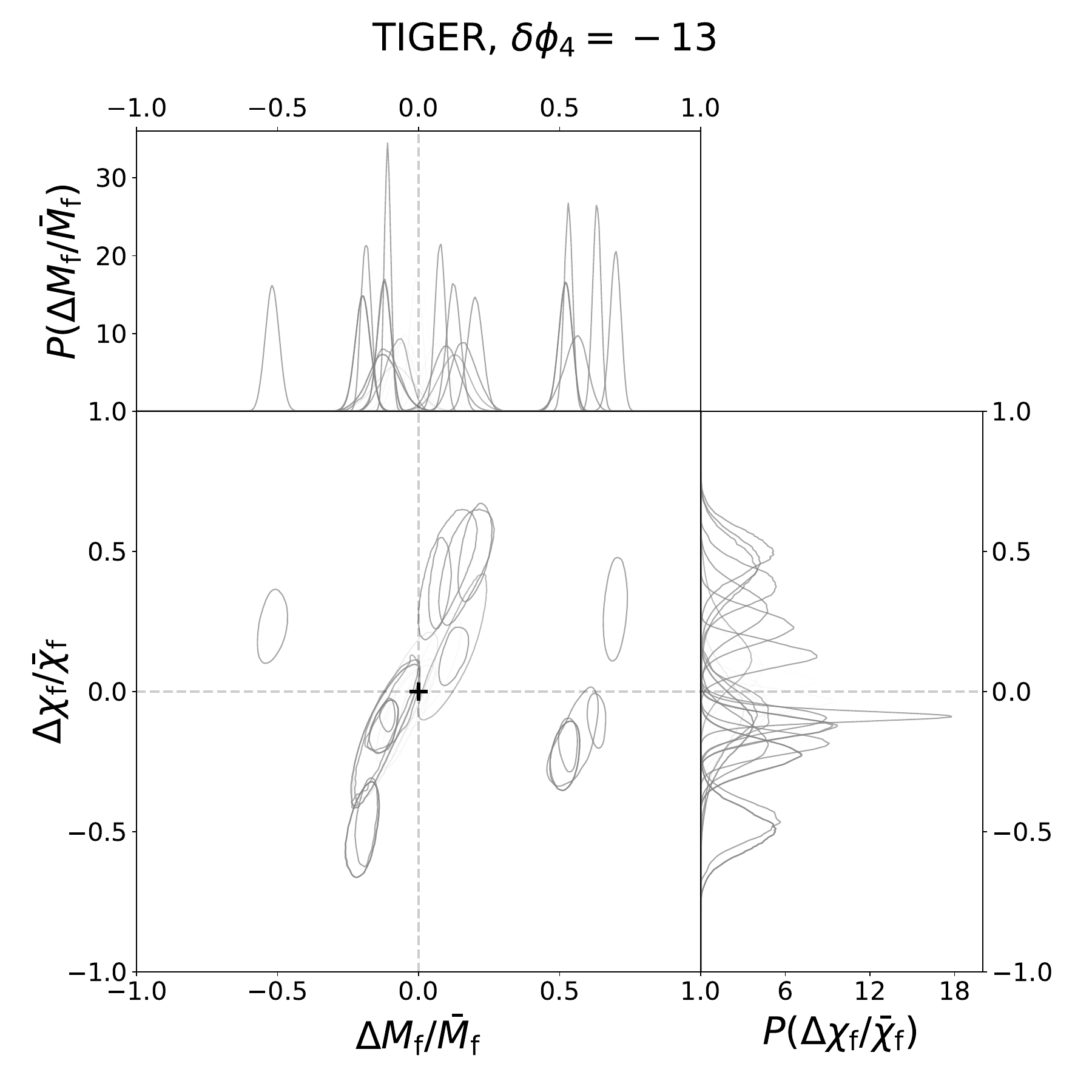}}
\qquad
\subfloat{\includegraphics[width=.5\textwidth]{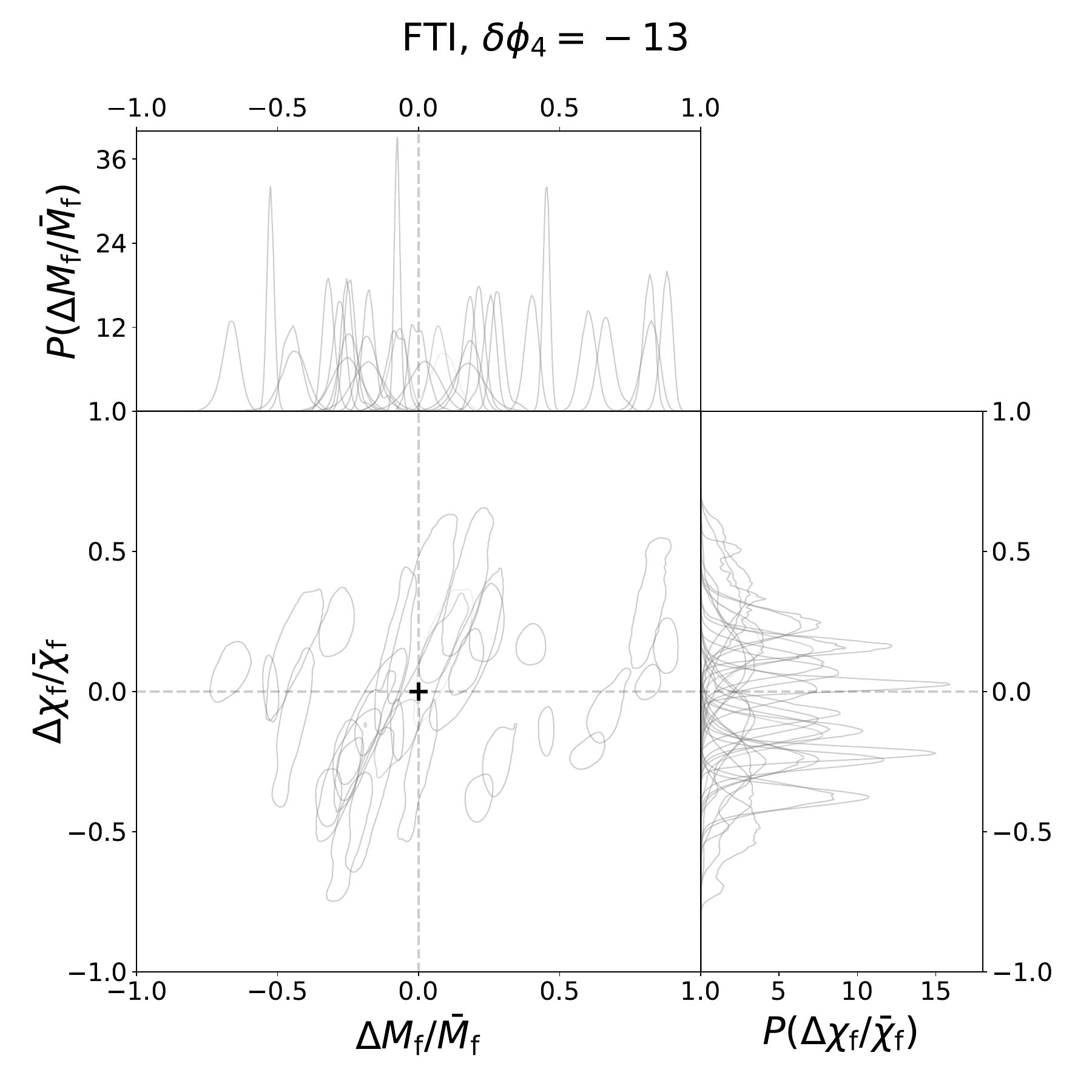}}

\caption[The results of meta IMRCT on the GW150914-like GR and larger GR deviation simulated observations]{\label{fig:GW150914-like_larger}The results of meta IMRCT on the GW150914-like GR simulated observations and those with the larger GR deviation presented
as 2d contour plots of the posteriors on the deviation parameters. The side panels shows the 1d histograms of the marginalized posteriors of $\Delta M_\text{f}/{\bar{M}_\text{f}}$ and $\Delta \chi_\text{f}/{\bar{\chi}_\text{f}}$. The gray contour delineates pairs of individual tests with color gradient varying based on the value of GR quantile (where darker gray denotes larger GR quantile and lighter gray denotes smaller GR quantile). We highlight in color the three cases with the largest meta IMRCT GR quantile where this is larger than the GR quantile from either of the tests of GR in the pair (or the single test of GR for pairs with GR PE) and also $> 90\%$. Specifically, we assign magenta color to the largest GR quantile, purple to the second-largest, and blue to the third-largest.}
\end{figure*}

\begin{figure*}[htb]
\centering
\subfloat{\includegraphics[width=.5\textwidth]{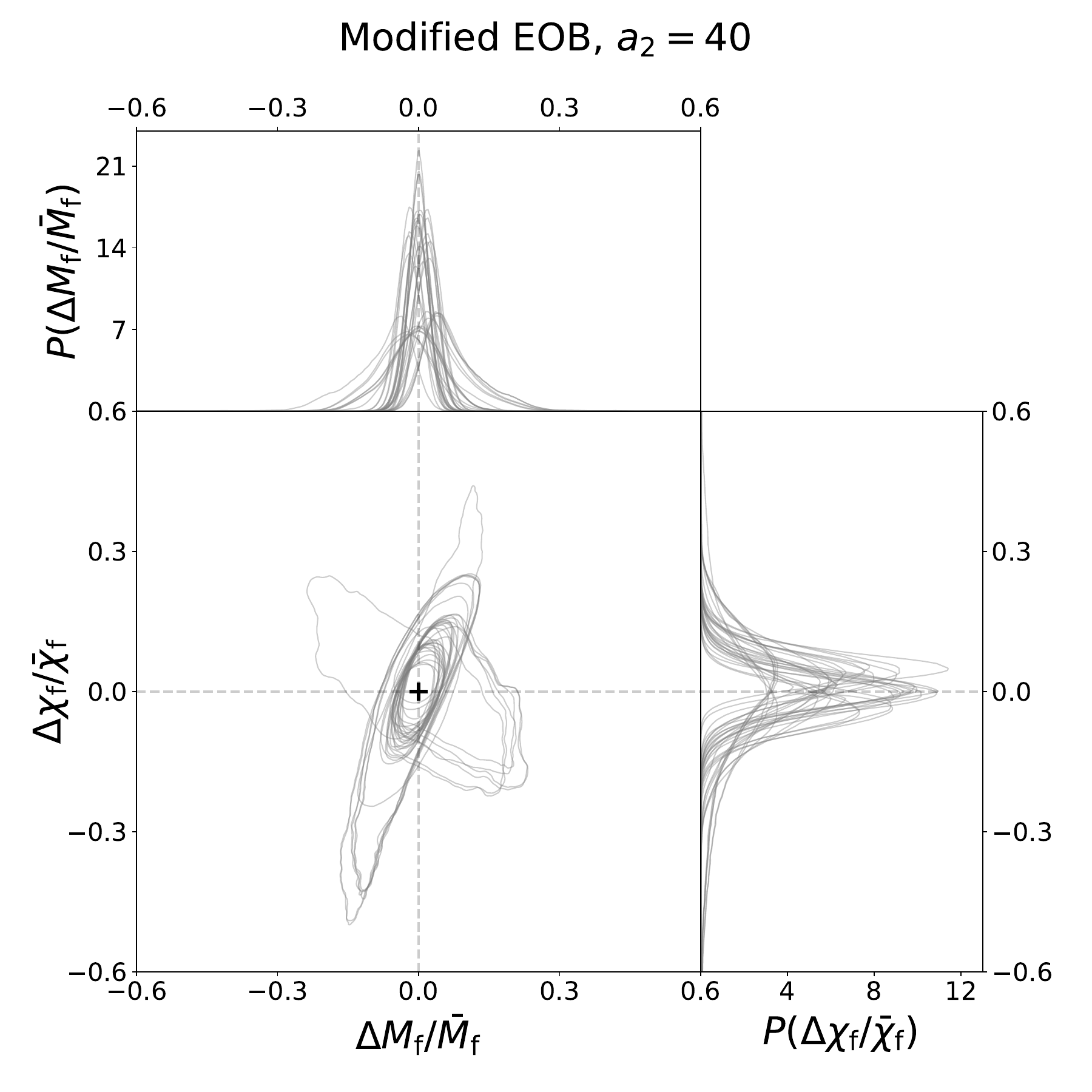}}
\qquad
\subfloat{\includegraphics[width=.5\textwidth]{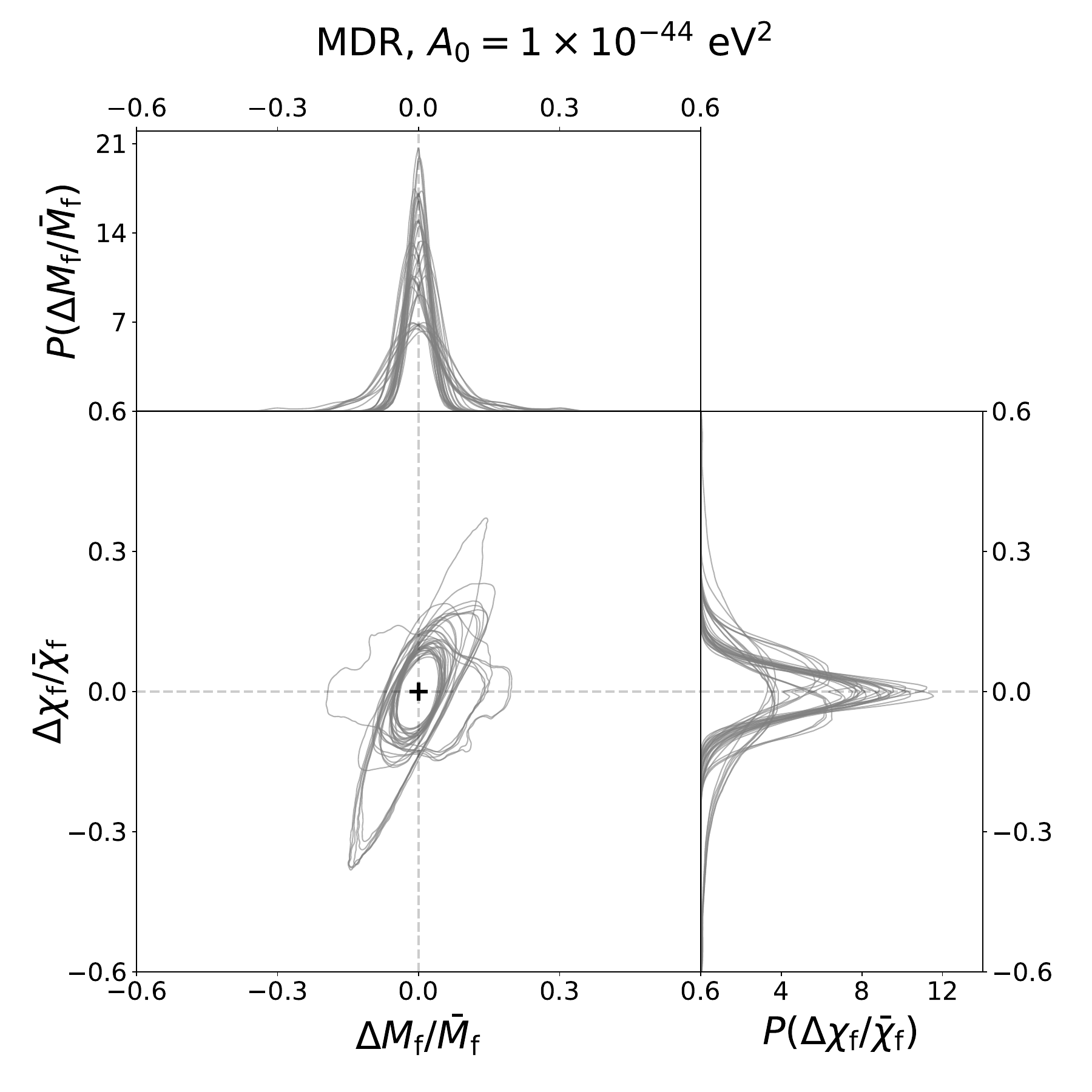}}
\\
\subfloat{\includegraphics[width=.5\textwidth]{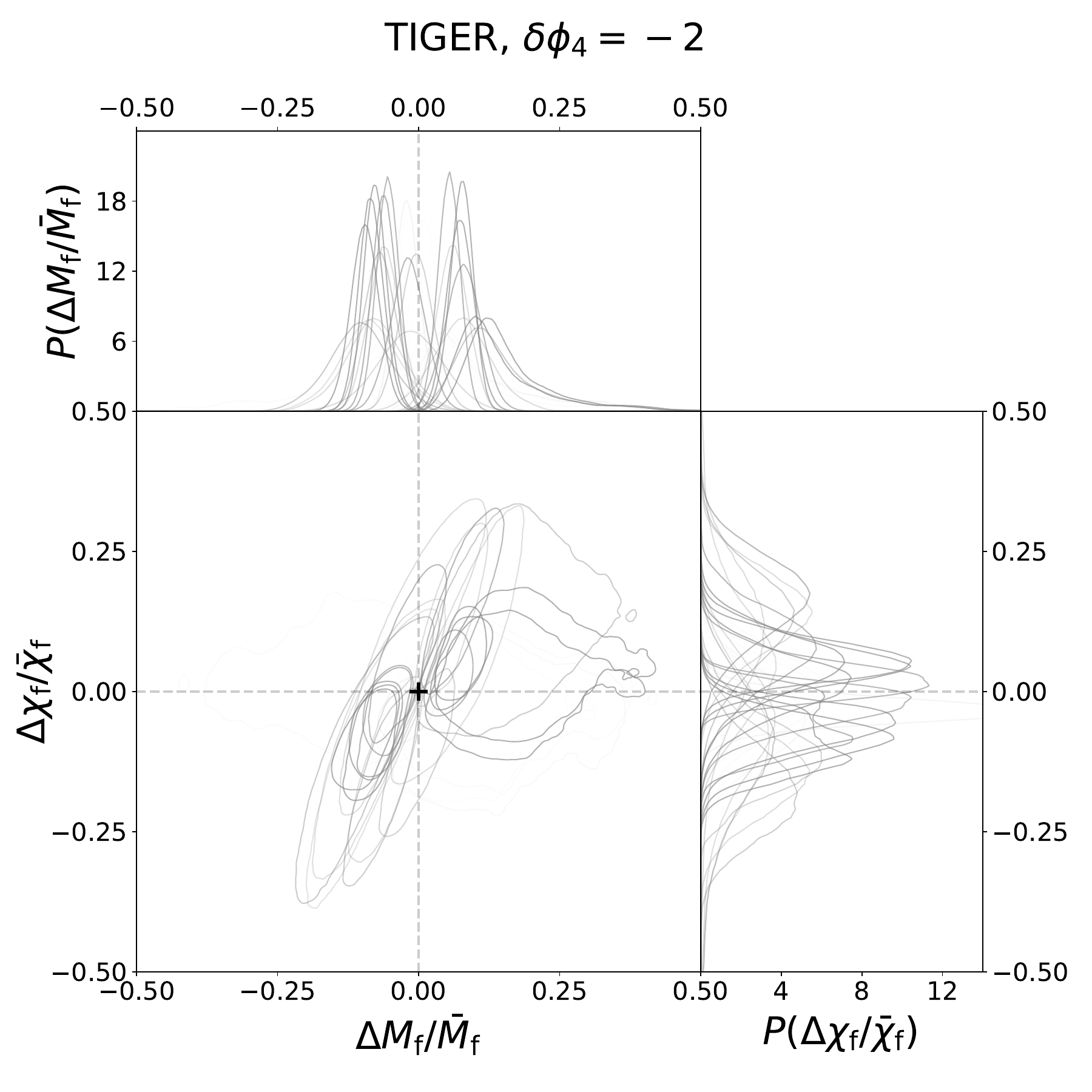}}
\qquad
 \subfloat{\includegraphics[width=.5\textwidth]{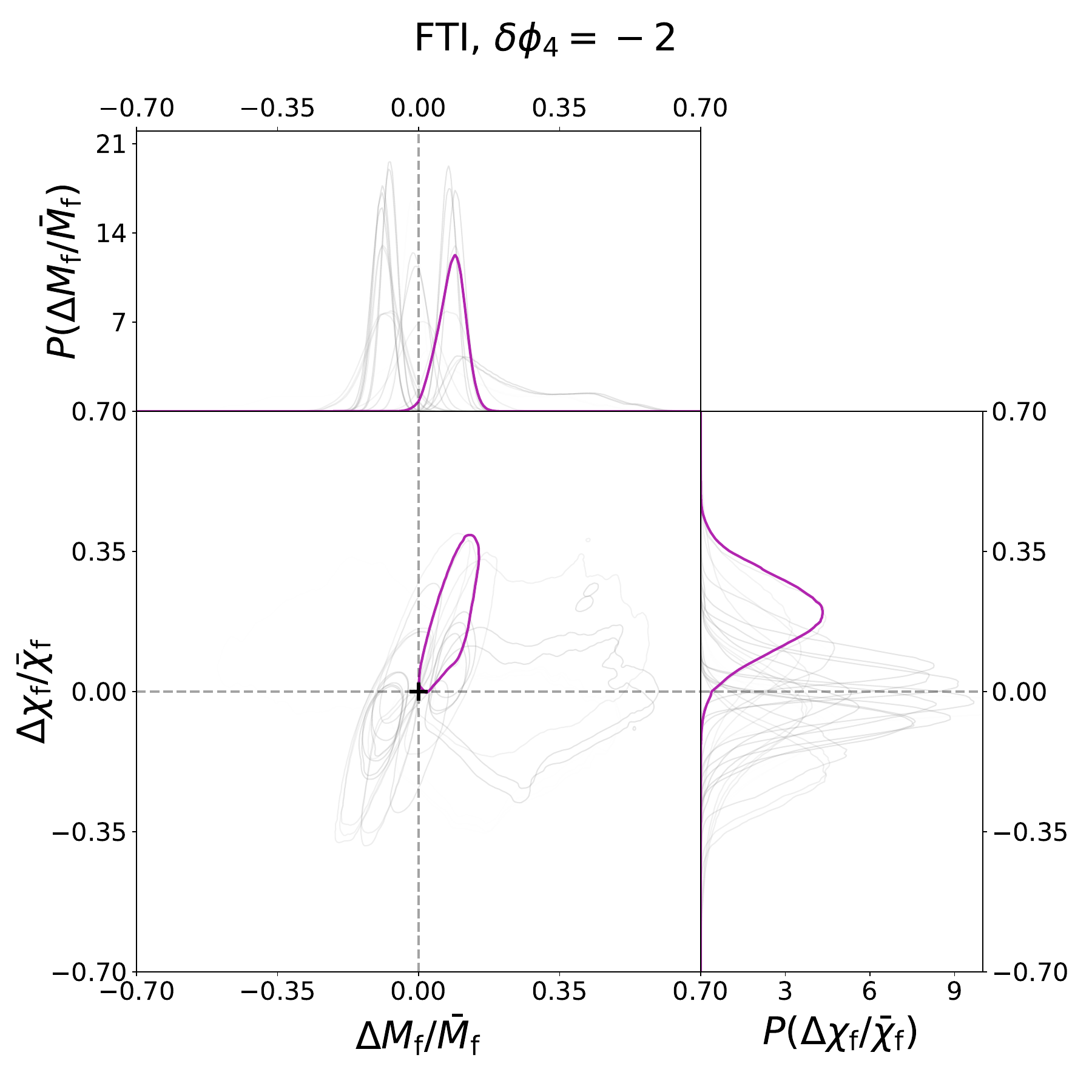}}

\caption[The results of meta IMRCT on the GW150914-like smaller GR deviation simulated observations]{\label{fig:GW150914-like_smaller}Similar to figure \ref{fig:GW150914-like_larger}, but for the GW150914-like smaller GR deviation simulated observations.}
\end{figure*}

\subsection{GW150914-like cases}
\label{ssec:GW150914}

We summarize the meta IMRCT results for the GW150914-like GR and non-GR simulated observations in table~\ref{tab:GW150914-like} as well as in figures~\ref{fig:GW150914-like_larger} (GR and larger GR deviations) and~\ref{fig:GW150914-like_smaller} (smaller GR deviations) \footnote{The sign of the deviation parameters depends on the ordering of the pair. We choose the following ordering for these plots and the similar ones below: GR PE, I, MR, $\delta\varphi_2^\text{TIGER}$, $\delta\varphi_4^\text{TIGER}$, $\delta\alpha_2$, $\delta\beta_2$, $\delta \varphi_2^\text{FTI}$, $\delta \varphi_4^\text{FTI}$, $A_0^{<0}$, $A_0^{>0}$, where not all of these are present for a given simulated observation.}. This table (and similar ones later) has a format where the different simulated observations are in separate rows, while the columns give the results of the different tests used in the application of the meta IMRCT as well as summaries of the meta IMRCT results, including various comparisons of the meta IMRCT GR quantiles with those of the individual tests, discussed below.

The meta IMRCT indeed finds that the GR simulated observations are consistent with GR, and recovers deviations from GR for many pairs for the larger GR deviations. In particular, the median GR quantile is $\geq 99.6\%$ for all the larger GR deviations except for the MDR simulated observation (though there are some pairs with considerably smaller GR quantiles for all simulated observations), while the median GR quantiles for the GR simulated observations are $< 4\%$. Indeed, the maximum GR quantiles for the GR simulated observations are only $68\%$ and $64\%$, respectively, for the EOB and Phenom versions, coming from the $(\delta\varphi_2^\text{TIGER},\delta\alpha_2)$ and $(\delta\varphi_2^\text{TIGER},\delta\beta_2)$ pairs, respectively. We also find that the combined $p$-values give good consistency with GR for the GR simulated observations and are smaller (i.e.\ there is a more significant GR deviation) when the GR quantile (and Gaussian sigma) and $N^{100\%}/N_\text{tot}$ values are larger, as expected. In particular, the most significant Simes $p$-value is $3.50\times 10^{-3}$ for the FTI simulated observation with the larger GR deviation. This case also gives the most significant Hommel $p$-value, of $1.45\times10^{-2}$, which, while a factor of $\sim 3$ less significant than the Simes result, is still an order of magnitude more significant than the best significance of $0.105$ that is obtainable with the classical Bonferroni method in this case, given that the individual pair $p$-values are no more significant than $3\times10^{-3}$, as discussed below. (A reminder that the Hommel method results are more conservative than the Simes method results, since we have not checked that the hypotheses necessary for the Simes method to give conservative results are satisfied by the meta IMRCT, since this does not seem to be straightforward, even though the Simes method is applicable to many situations encountered in practice.)

For a few of the simulated observations with larger GR deviations, the meta IMRCT gives a GR quantile $\geq 90\%$ and this GR quantile is larger than that from either of the individual tests (this is the number $N_\text{mIMRCT}^{\text{grtr}, 90\%}$ in the table) \footnote{There are additional cases where the meta IMRCT gives a larger GR quantile than do the individual tests, but this GR quantile is $< 90\%$, and thus there is not a significant GR deviation identified. We thus give such results in the tables with $N_\text{mIMRCT}^\text{grtr}$ but do not comment on them in detail.}. Here, to make this comparison, we have converted the 1d GR quantiles $Q_\text{1d}$ for the individual tests to the equivalent of the 2d GR quantile, given by $|2Q_\text{1d} - 1|$ (so that $0$ indicates agreement with GR). Moreover, here and subsequently, we round (2d or equivalent) GR quantiles above $99.7\%$ to $100\%$, except when computing the combined $p$-value, we round them to $99.7\%$, to be conservative. This lower bound arises because larger GR quantiles cannot be stated with certainty from the $\sim 10^4$ posterior samples we have, as discussed in~\cite{Narayan:2023vhm}. We thus also quote in the tables (as $N^{100\%}$) the number of pairs for which the meta IMRCT and at least one of the individual tests both give a GR quantile of $100\%$. For the modified EOB simulated observation with the larger GR deviation, the pairs of $(\delta \varphi_2^\text{TIGER}, \delta \alpha_2)$ and $(\delta \varphi_2^\text{TIGER}, \delta \beta_2)$ gave GR quantiles of $100\%$ and $99.7\%$, respectively, which are greater than the corresponding individual GR quantiles, which are $99.6\%$, $95.6\%$ and $27.8\%$ for $\delta \varphi_2^\text{TIGER}$, $\delta \alpha_2$ and $\delta \beta_2$, respectively. We also find that there are $25$ pairs (close to half of the total of $54$ pairs) where the meta IMRCT GR quantile and the larger of the two individual GR quantiles both round to $100\%$. In such cases a significantly larger number of samples would be necessary to determine whether the meta IMRCT identifies the GR deviation with higher significance than the individual tests do.

For the MDR simulated observation with the larger GR deviation, the meta IMRCT only gives a median GR quantile of $71.5\%$, but there is still one pair for which the meta IMRCT gives a larger GR quantile than individual tests: $(\delta \varphi_2^\text{FTI},\text{I})$ gives a meta IMRCT GR quantile of $91.6\%$, while the individual GR quantiles are $84.4\%$ and $90.1\%$ for $\delta \varphi_2^\text{FTI}$ and the standard IMRCT, respectively. For the TIGER and FTI simulated observations with the larger GR deviation, there are no cases where the meta IMRCT gives a larger GR quantile than the individual tests do, though there are many cases where the meta IMRCT and at least one of the individual tests both identify these very significant GR deviations with a GR quantile of $100\%$. Indeed, this is the case for $17$ ($49\%$) of the pairs for the TIGER simulated observation and $30$ ($86\%$) of the pairs for the FTI simulated observation.

For the GW150914-like cases with smaller GR deviations, the meta IMRCT does not recover any significant deviation from GR in the modified EOB or massive graviton simulated observations, giving maximum GR quantiles of $75.6\%$, though the largest GR quantile from the individual tests in those cases is only $90.7\%$, for the MDR analysis of the modified EOB simulated observation. However, the meta IMRCT recovers a slightly larger GR quantile than the larger GR quantile for the individual tests for the $(\delta \beta_2, \delta \alpha_2)$ pair and the FTI simulated observation, where the pair gives a GR quantile of $97.81\%$, while the individual tests give GR quantiles of $97.77\%$ and $12.69\%$, respectively. There are also $10$ pairs for which the meta IMRCT GR quantile and the larger of the two individual GR quantiles both round to $100\%$ for the FTI simulated observation, as well as $4$ such pairs for the TIGER simulated observation, though there are no pairs for which the meta IMRCT gives a larger GR quantile.
\subsection{GW170608-like cases}
\label{ssec:GW170608}

\begin{table*}[h!]

\caption[Summary of meta IMRCT on GW170608-like simulated observations]{\label{tab:GW170608-like}The analogue of table~\ref{tab:GW150914-like} for the GW170608-like simulated observations.}
\scalebox{0.7}{
\begin{tabular}{*{25}{c}}
\hline\hline
\\
\multirow{2}{*}{Simul.\ Obs.\ } & \multirow{2}{*}{GR PE} & \multirow{2}{*}{IMRCT }& \multirow{2}{*}{TIGER} & \multirow{2}{*}{FTI} & \multirow{2}{*}{MDR} & \multirow{2}{*}{GR Quantile (\%)} & \multirow{2}{*}{Gaussian $\sigma$} & Combined $p$-value&\multirow{2}{*}{$N^{100\%}/N_\text{mIMRCT}^{\text{grtr}, 90\%}/N_\text{mIMRCT}^\text{grtr}/N_\text{tot}$ }\\
&   &  & & & &  &  &$p_\text{S}, p_\text{H}$&\\
\\
\hline

EOB, GR & \checkmark& $\times$ & $\delta \varphi_4,\delta \alpha_2,\delta \beta_2$&   $ \delta \varphi_4$ &$\tilde{A}_0$  & $16^{+10.6}_{-15.4}$ & $0.2^{+0.1}_{-0.2}$ & $0.996, 1$ &$0/0/3/21$\\[1pt]
\\
Phenom, GR & \checkmark& $\times$ & $\delta \varphi_4,\delta \alpha_2,\delta \beta_2$&  $ \delta \varphi_4$   & $\tilde{A}_0$   & $17.9^{+10.9}_{-17.2}$ &$0.2^{+0.1}_{-0.2}$  & $0.994, 1$ &$0/0/10/21$\\[1pt]
\\
modified EOB & \checkmark& $\times$& $\delta \varphi_4,\delta \alpha_2,\delta \beta_2$&$ \delta \varphi_4$ &$\tilde{A}_0$  & $27.2^{+18.6}_{-26.1}$ & $0.3^{+0.3}_{-0.3}$& $0.997, 1$ &$0/0/6/21$\\[1pt]
\\
MDR & \checkmark& $\times$ & $\delta \varphi_4,\delta \alpha_2,\delta \beta_2$& $ \delta \varphi_4$& $\tilde{A}_0$ & $63.7^{+36.3}_{-62.8}$ & $0.9^{+2.1}_{-0.9}$ &$1.26\times 10^{-2}, 4.59\times 10^{-2}$ &$5/0/4/21$\\[1pt]
\\
TIGER & \checkmark& $\times$ & $\delta \varphi_4,\delta \alpha_2,\delta \beta_2$& $ \delta \varphi_4$&$\tilde{A}_0$  & $72.4^{+17.3}_{-28.1}$ & $0.1^{+0.5}_{-0.5}$& $0.460, 1$ &$0/0/3/21$\\[1pt]
\\
FTI & \checkmark& $\times$ & $\delta \varphi_4,\delta \alpha_2,\delta \beta_2$& $ \delta \varphi_4$&$\tilde{A}_0$  & $55.7^{+24.6}_{-55.5}$ & $0.8^{+0.5}_{-0.8}$ & $0.748, 1$ &$0/0/6/21$\\[1pt]

\hline\hline
\end{tabular}
}
\label{tab:Table 5.2}
\end{table*}

\begin{figure}[htb]
\centering

 \subfloat{\includegraphics[width=.5\textwidth]{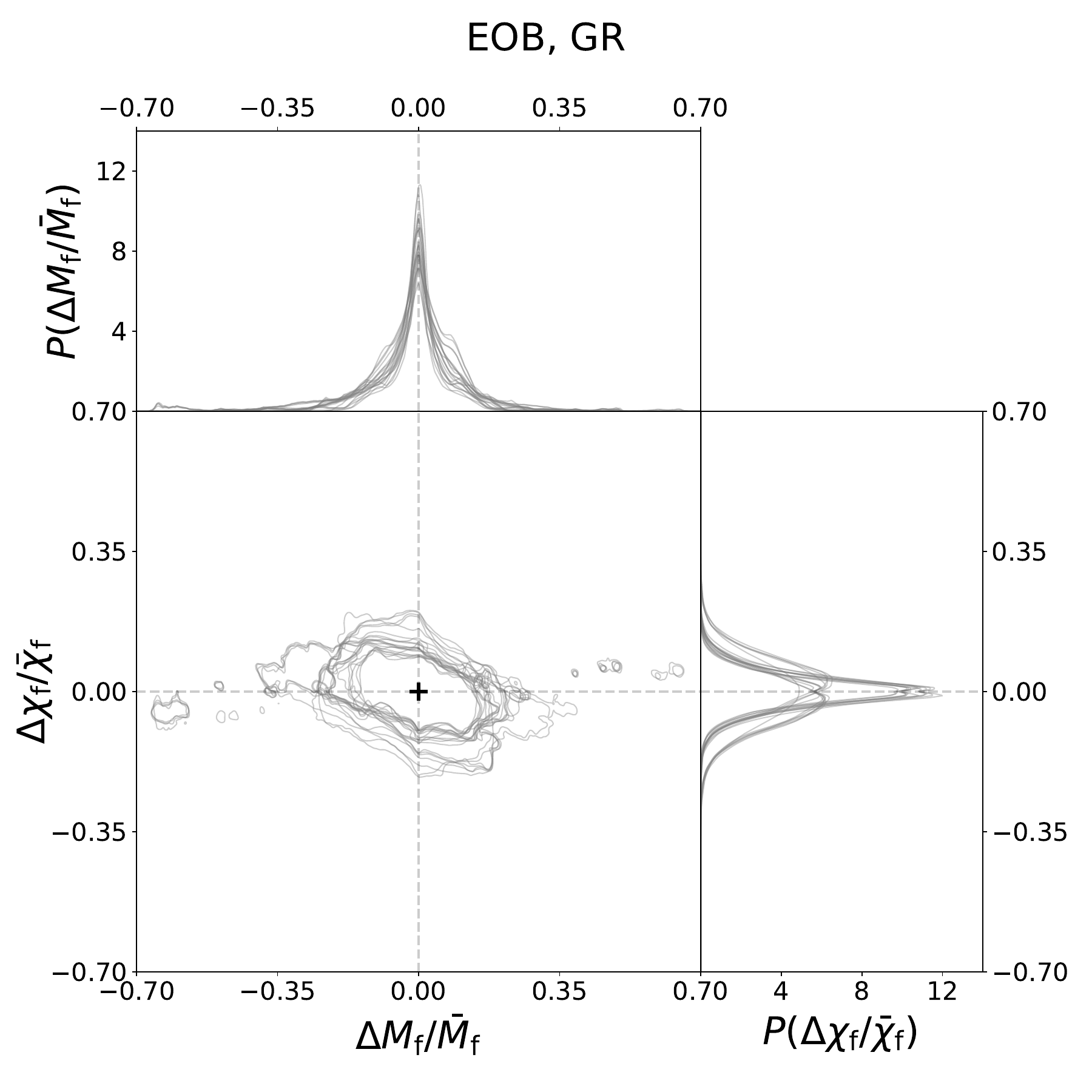}}
  \qquad
 \subfloat{\includegraphics[width=.5\textwidth]{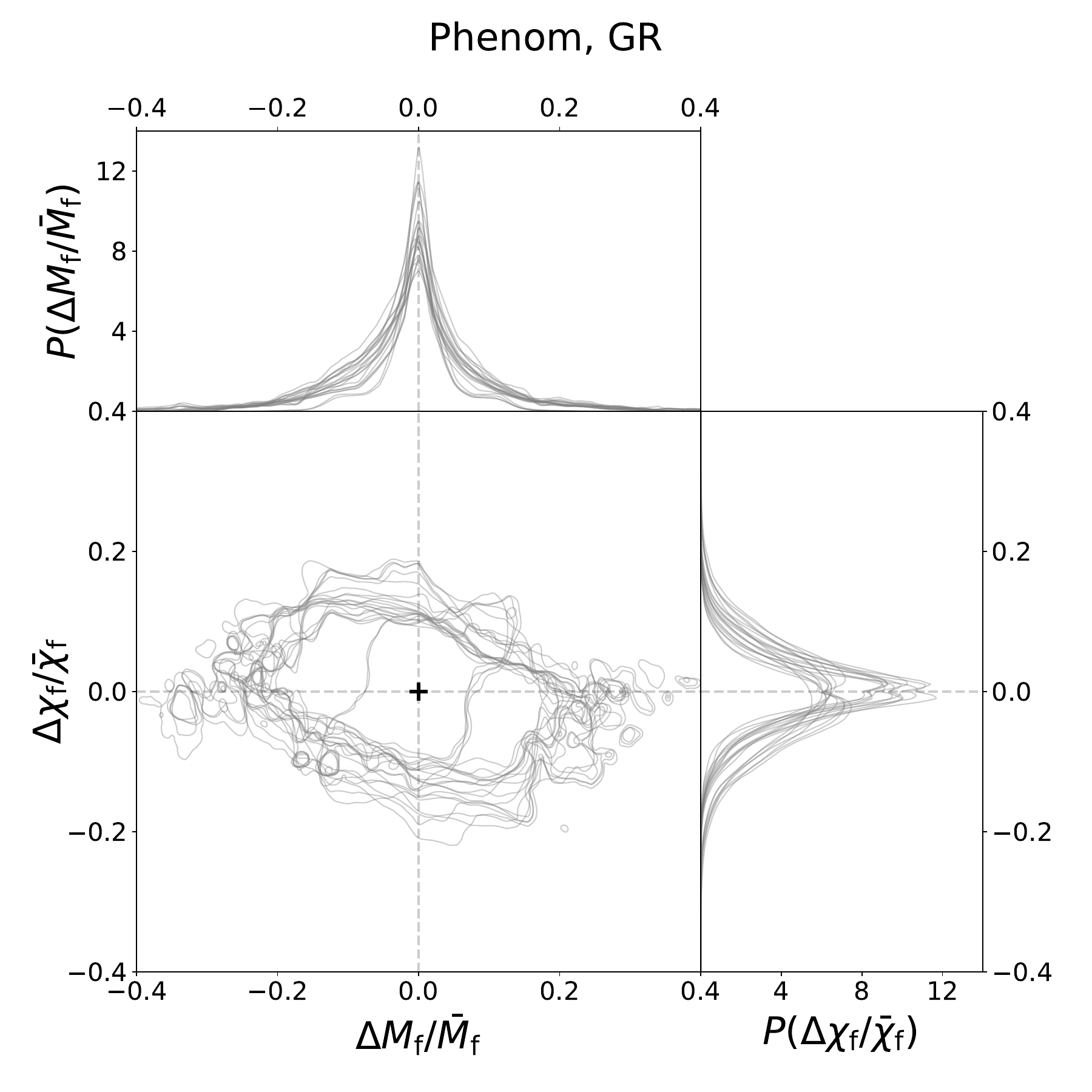}}
\\
\subfloat{\includegraphics[width=.5\textwidth]{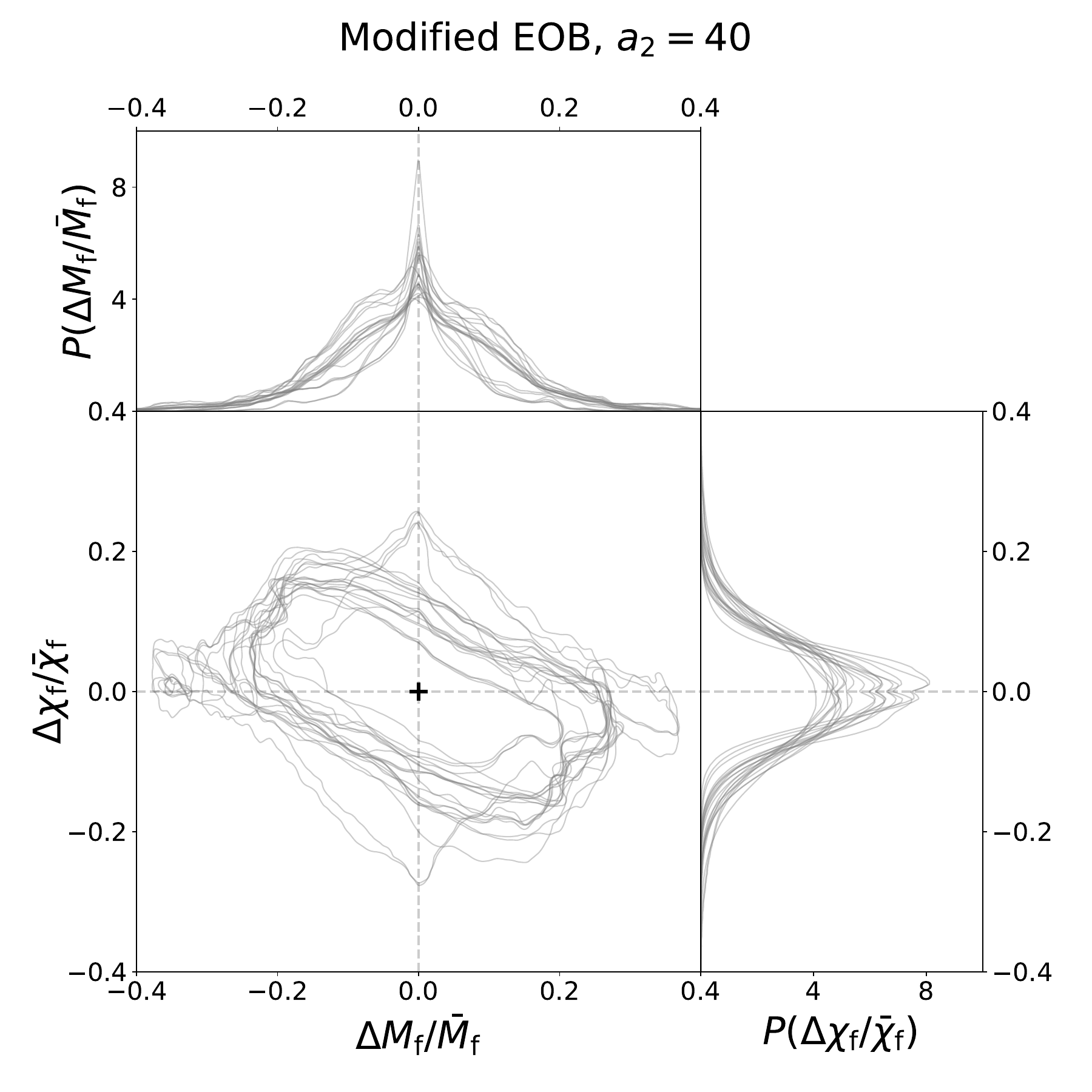}}
\qquad
\subfloat{\includegraphics[width=.5\textwidth]{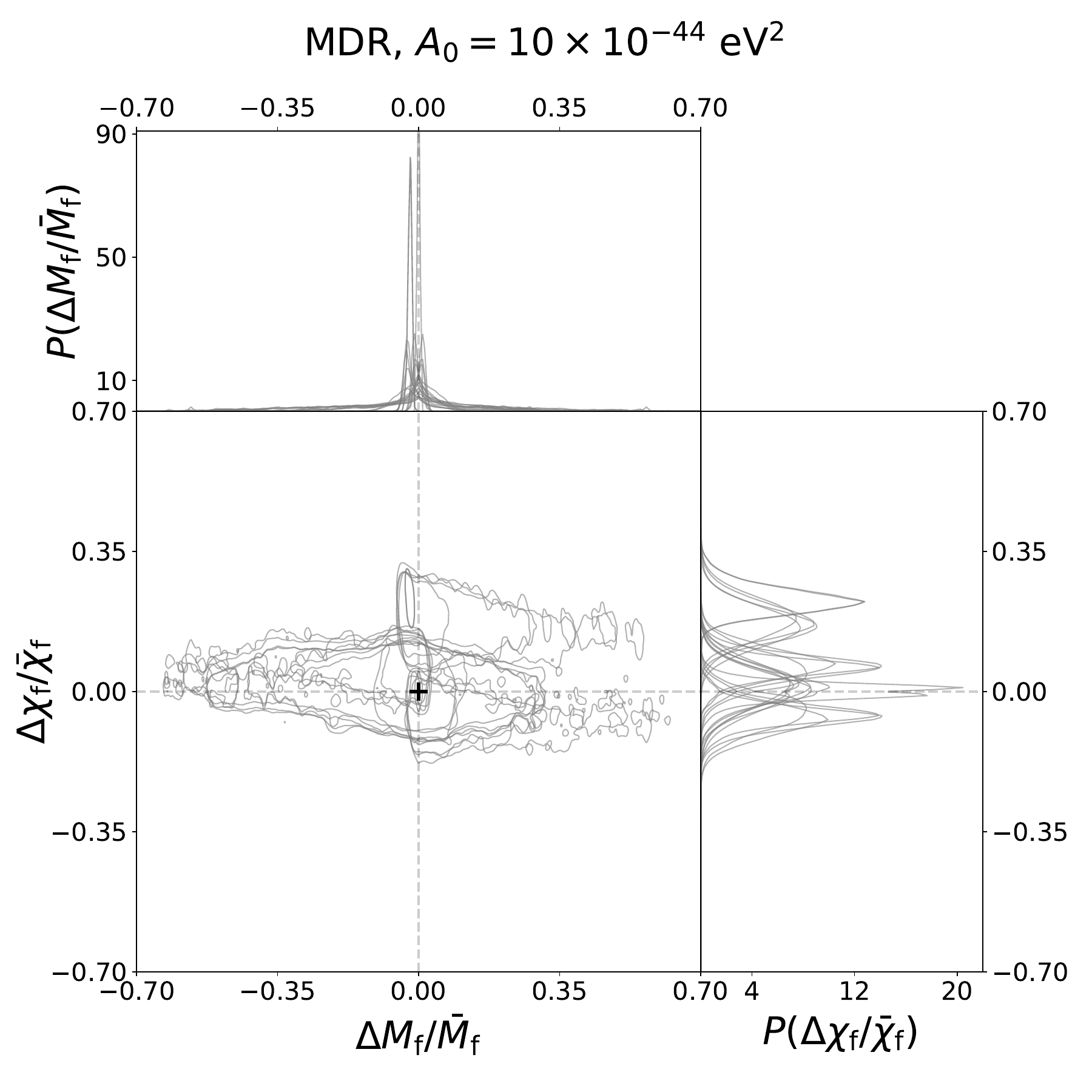}}
\\
\subfloat{\includegraphics[width=.5\textwidth]{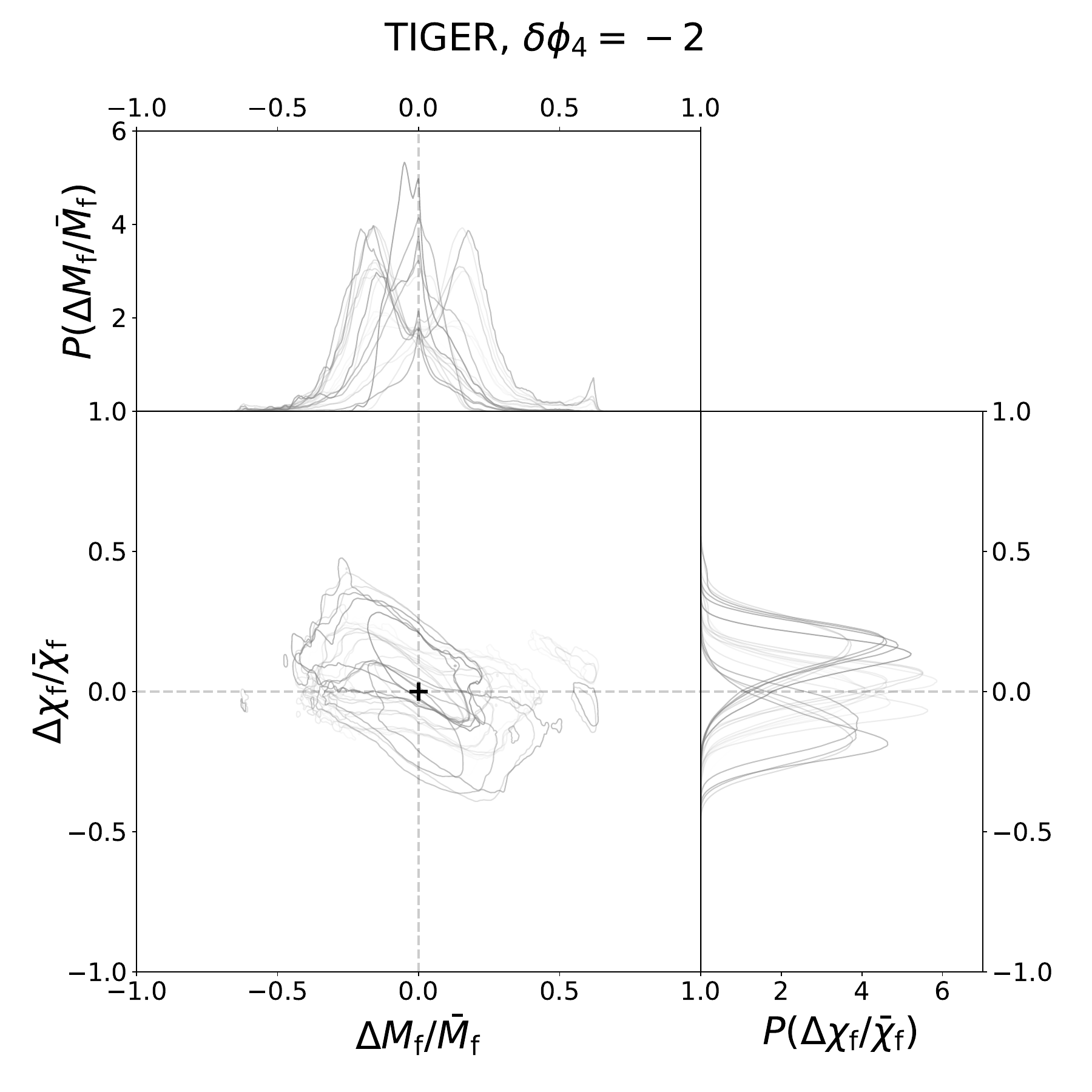}}
\qquad
\subfloat{\includegraphics[width=.5\textwidth]{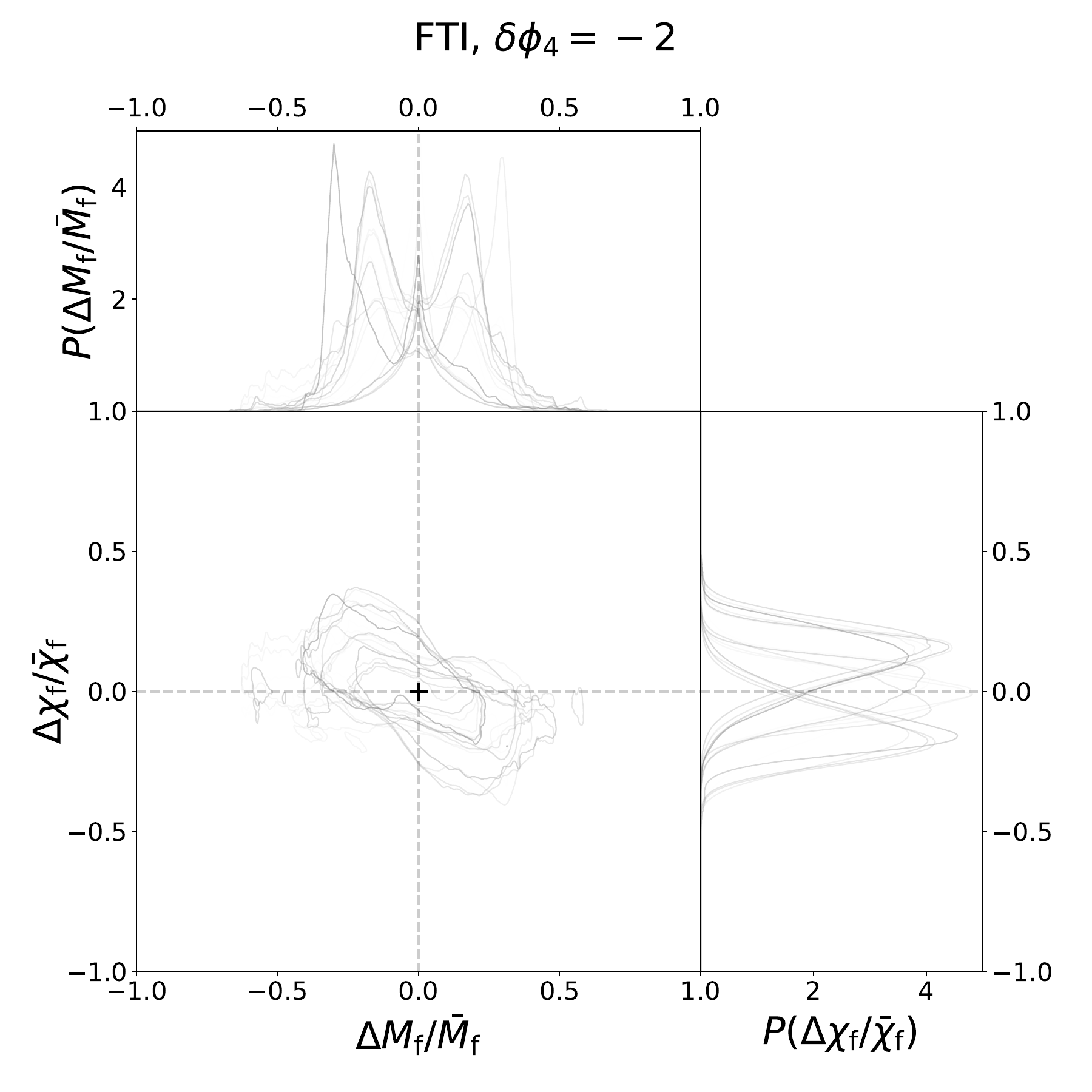}}

\caption[The results of meta IMRCT on the GW170608-like simulated observations]{\label{fig:GW170608-like}Similar to figure \ref{fig:GW150914-like_larger} for the GW170608-like smaller GR deviation simulated observations.}
\end{figure}

The GW170608-like simulated observations are for GR signals and non-GR signals with smaller GR deviations. We show the results for these simulated observations in table~\ref{tab:GW170608-like} and figure~\ref{fig:GW170608-like}. The meta IMRCT finds that the GR signals are indeed consistent with GR, with a largest GR quantile of $30.4\%$ [for the $(\delta \varphi_4^\text{TIGER}, \delta \alpha_2)$ pair and the Phenom simulated observation]. For the modified EOB simulated observation, none of the individual tests find a significant GR deviation (largest GR quantile of $71.7\%$ for the MDR analysis), and the meta IMRCT finds even smaller GR quantiles [the largest is $49.4\%$ for the $(\delta \varphi_4^\text{TIGER}, \delta \alpha_2)$ pair]. However, for the MDR simulated observation, the MDR test finds a significant GR deviation, with a GR quantile of $100\%$, though the maximum GR quantile for the other analyses is only $52.5\%$ for $\delta\beta_2$. The meta IMRCT obtains GR quantiles of $100\%$ for all pairs involving $A_0^{>0}$ (the sign of $A_0$ given in the simulated observation), except for $(\delta \varphi_4^\text{FTI}, A_0^{>0})$, where the GR quantile is $99.1\%$. The largest GR quantile for the pairs not involving $A_0^{>0}$ is $89.3\%$ for the $(\text{GR PE}, \delta\alpha_2)$ pair; the $\delta\alpha_2$ GR quantile is $20.7\%$. The combined $p$-value is also the smallest (most significant) for the MDR simulated observation, though not as significant as the largest significances found for the GW150914-like simulated observations.

For the TIGER simulated observation, the $\delta\varphi_4^\text{TIGER}$, $\delta\varphi_4^\text{FTI}$ and $\delta\beta_2$ analyses find GR quantiles $> 90\%$, though the meta IMRCT only gives a GR quantile $> 90\%$ for the $(\delta\varphi_4^\text{TIGER},\delta\alpha_2)$ pair, where the individual GR quantiles are $96.6\%$ and $34.6\%$, respectively. For the FTI simulated observation, the maximum GR quantile for the individual tests is $87.9\%$ for $\delta\varphi_4^\text{FTI}$. The maximum GR quantile from the meta IMRCT is $87.0\%$ for the $(\delta\varphi_4^\text{TIGER},\delta\alpha_2)$ pair, where the individual GR quantiles are $79.9\%$ and $29.7\%$, respectively.

\subsection{Quasicircular and eccentric numerical relativity waveforms}
\label{ssec:ecc and quasi}

\begin{figure}[htb]

\centering

\subfloat{\includegraphics[width=.5\textwidth]{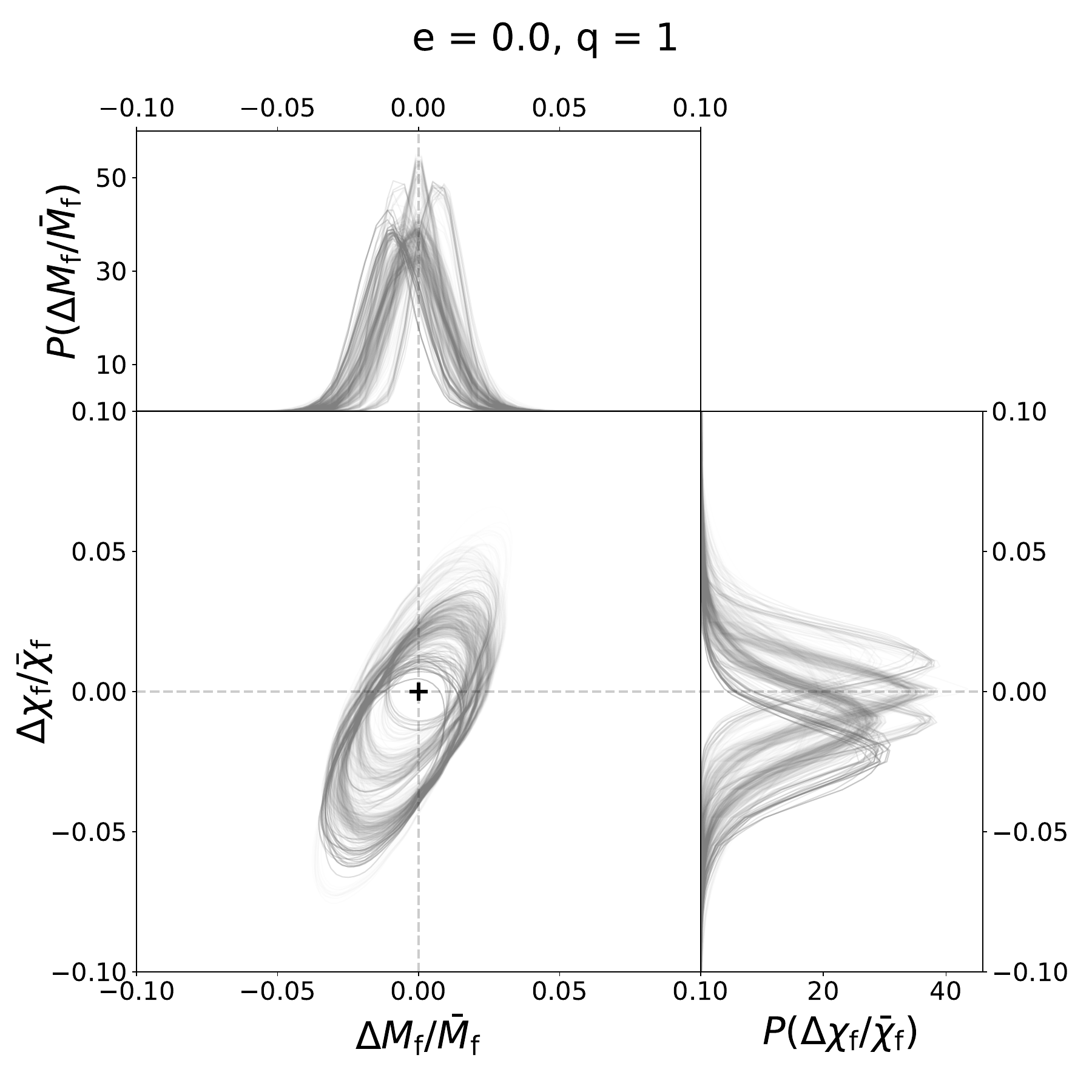}}
\\
\subfloat{\includegraphics[width=.5\textwidth]{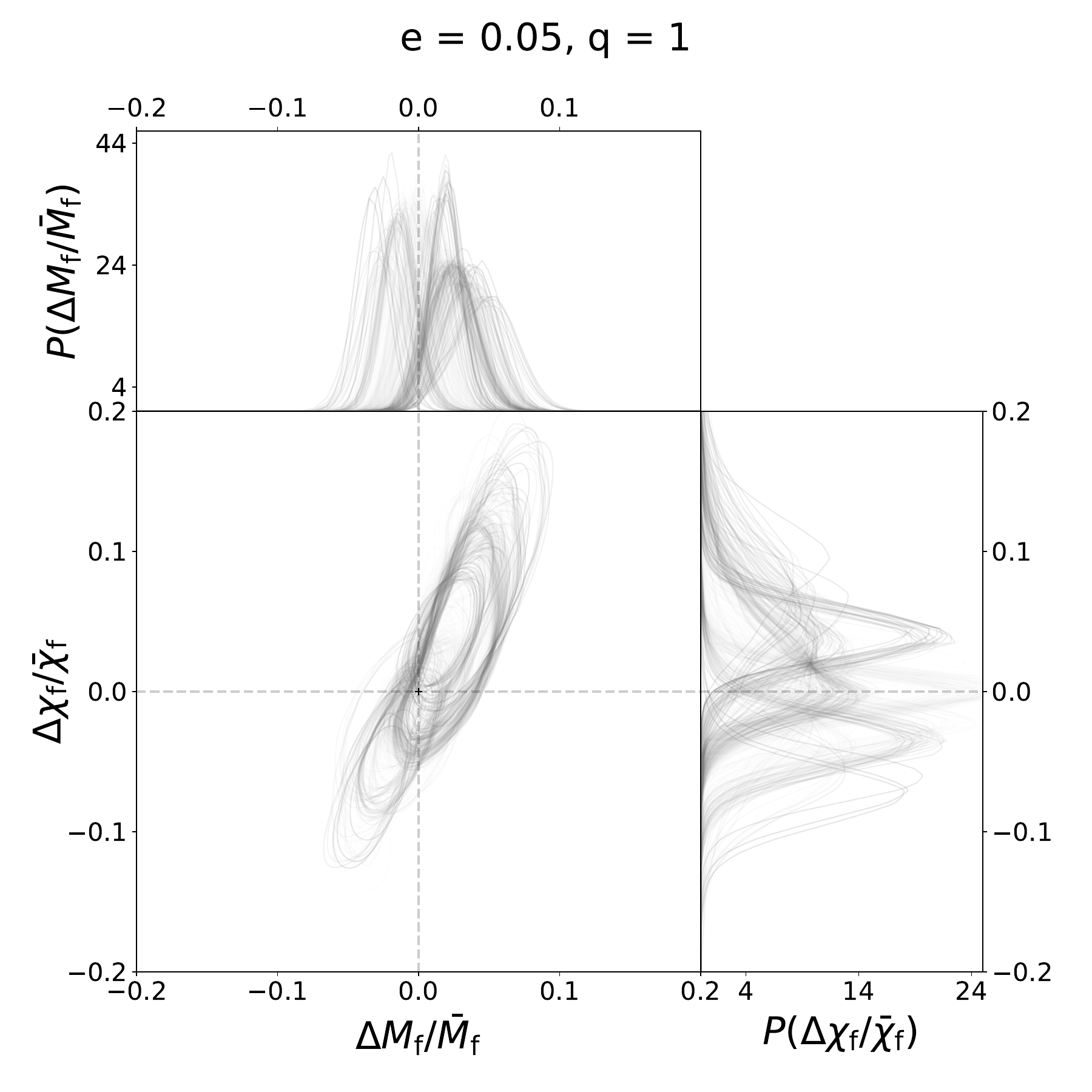}}
\qquad
\subfloat{\includegraphics[width=.5\textwidth]{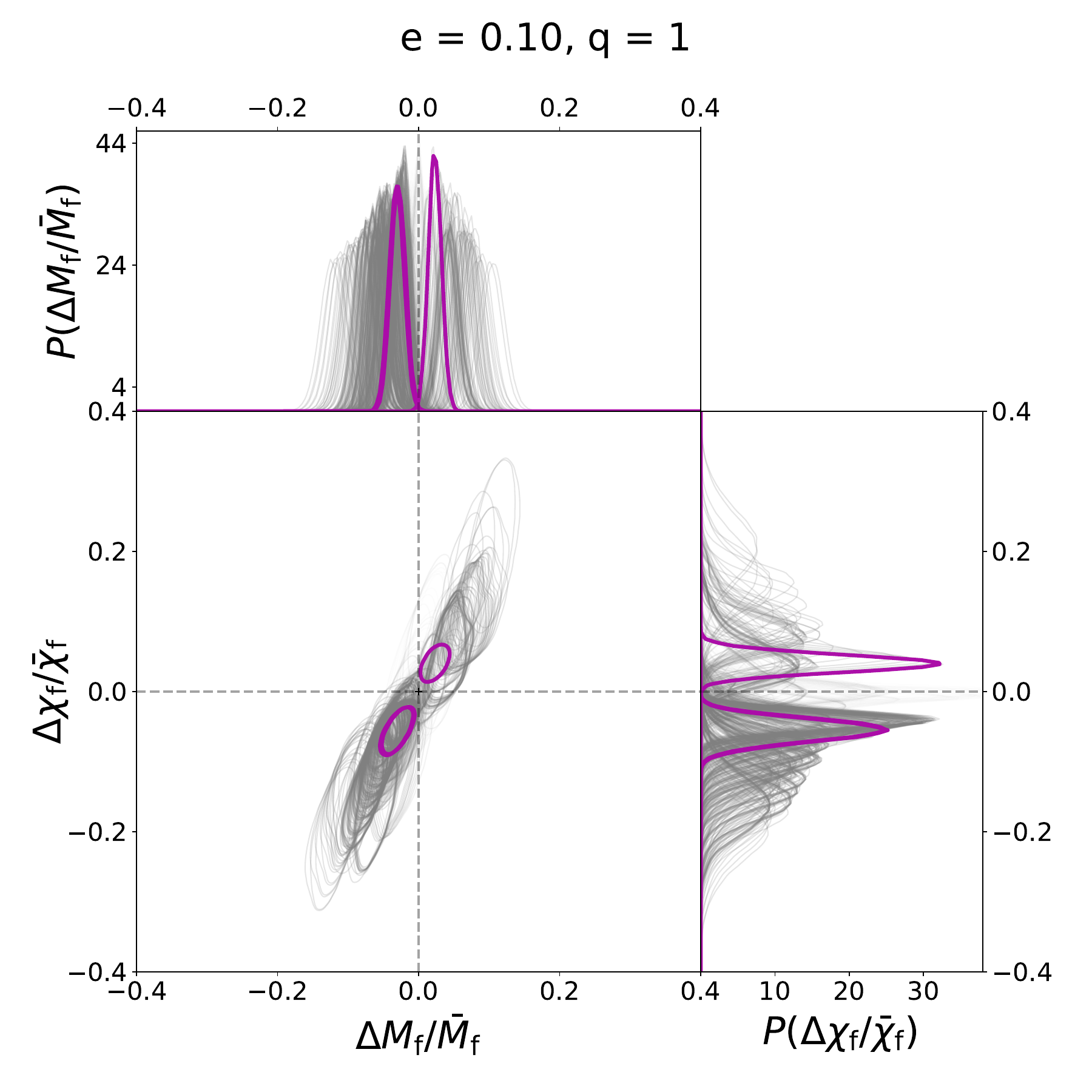}}

\caption[ The results of meta IMRCT on the quasicircular and eccentric NR simulations for $q = 1$]{\label{fig:figure 5.4}The results of meta IMRCT on the numerical relativity quasicircular and eccentric simulated observations with $q = 1$ presented as 2d contour plots of the posteriors on the deviation parameters. These are plotted the same way as in figure~\ref{fig:GW150914-like_larger}.}

\end{figure}

\begin{figure}[htb]

\centering

\subfloat{\includegraphics[width=.5\textwidth]{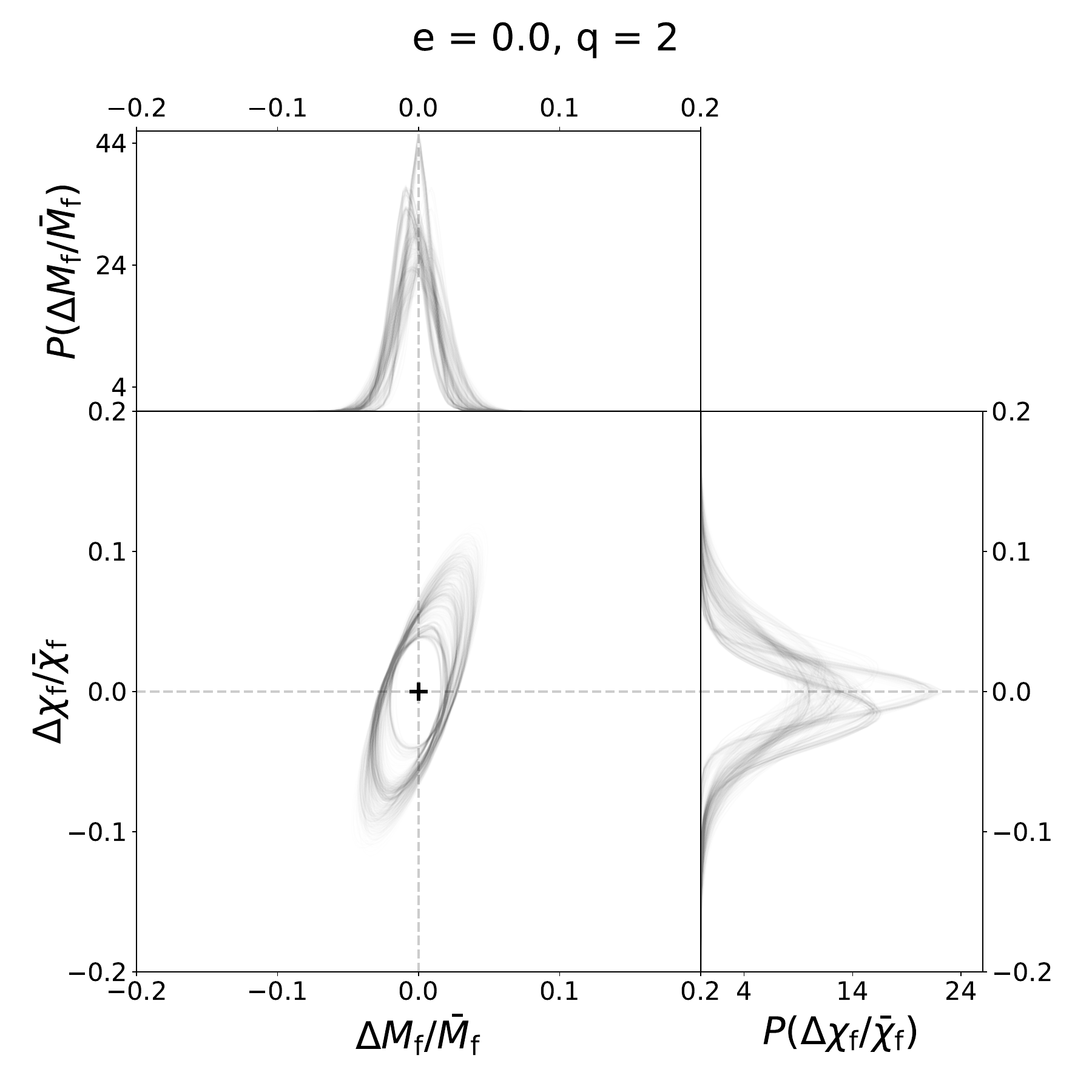}}
\\
\subfloat{\includegraphics[width=.5\textwidth]{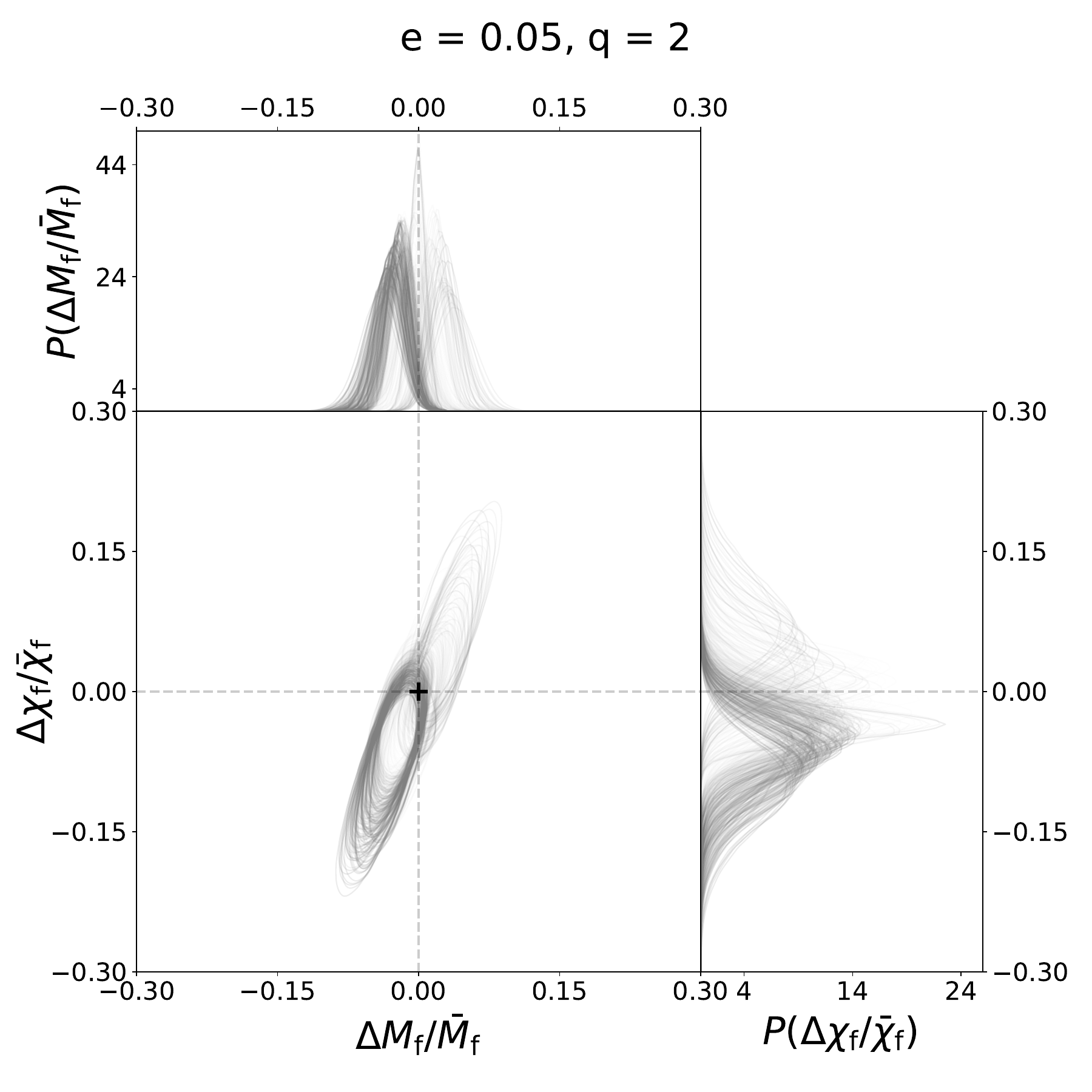}}
\qquad
\subfloat{\includegraphics[width=.5\textwidth]{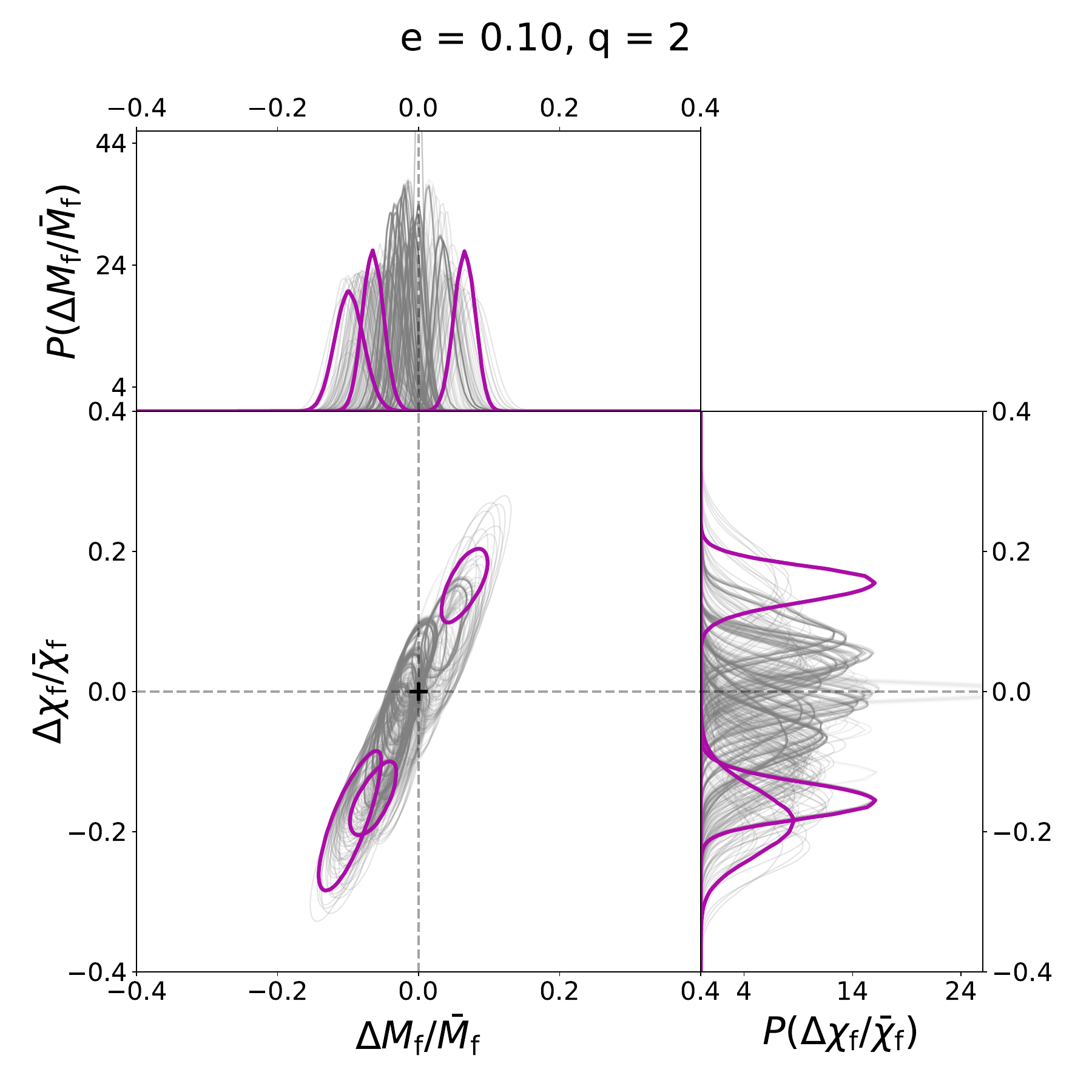}}

\caption[]{\label{fig:figure 5.5} Similar to figure \ref{fig:figure 5.4} for the $q = 2$ numerical relativity quasicircular and eccentric simulated observations.}

\end{figure}

\begin{figure}[htb]

\centering
\subfloat{\includegraphics[width=.5\textwidth]{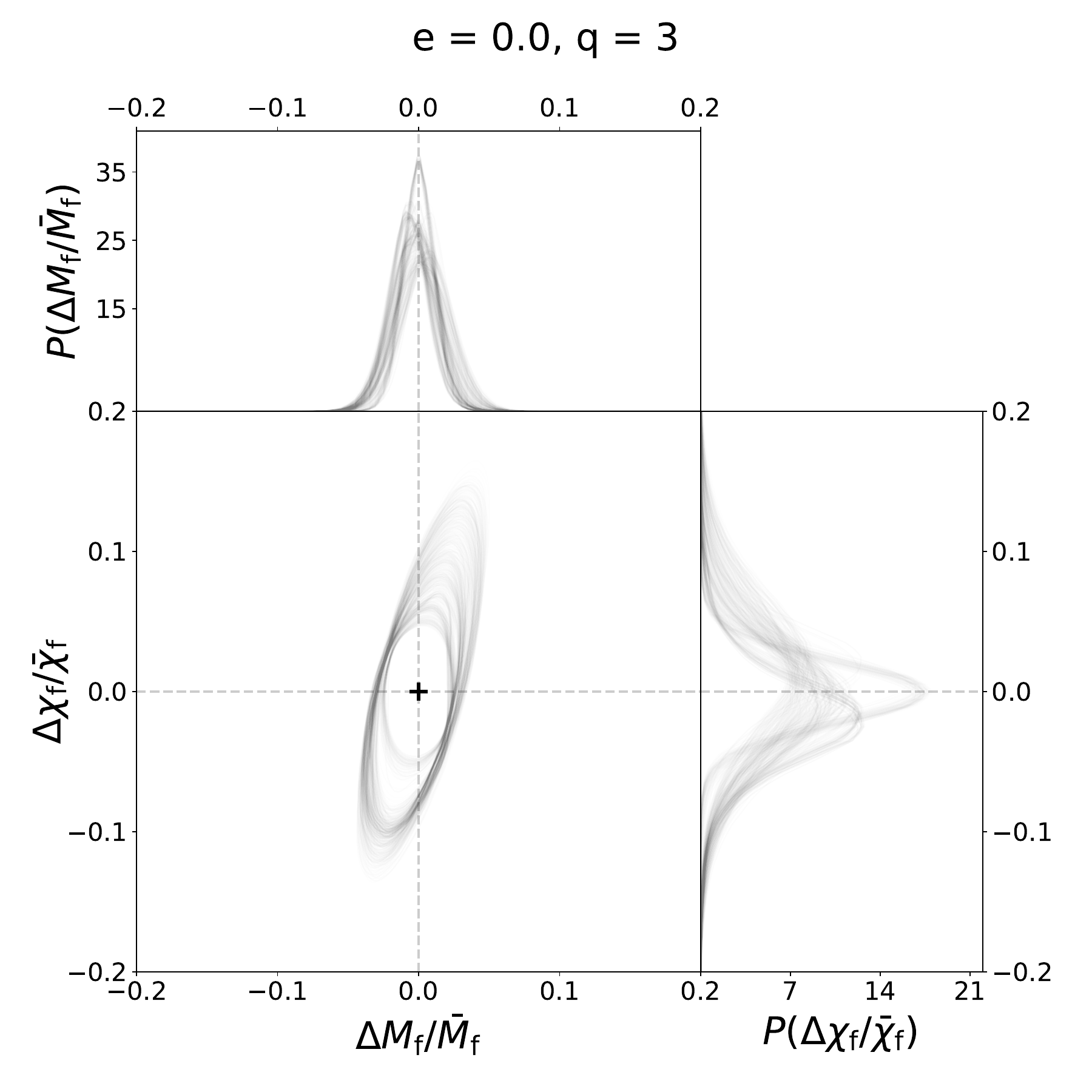}}
\\
\subfloat{\includegraphics[width=.5\textwidth]{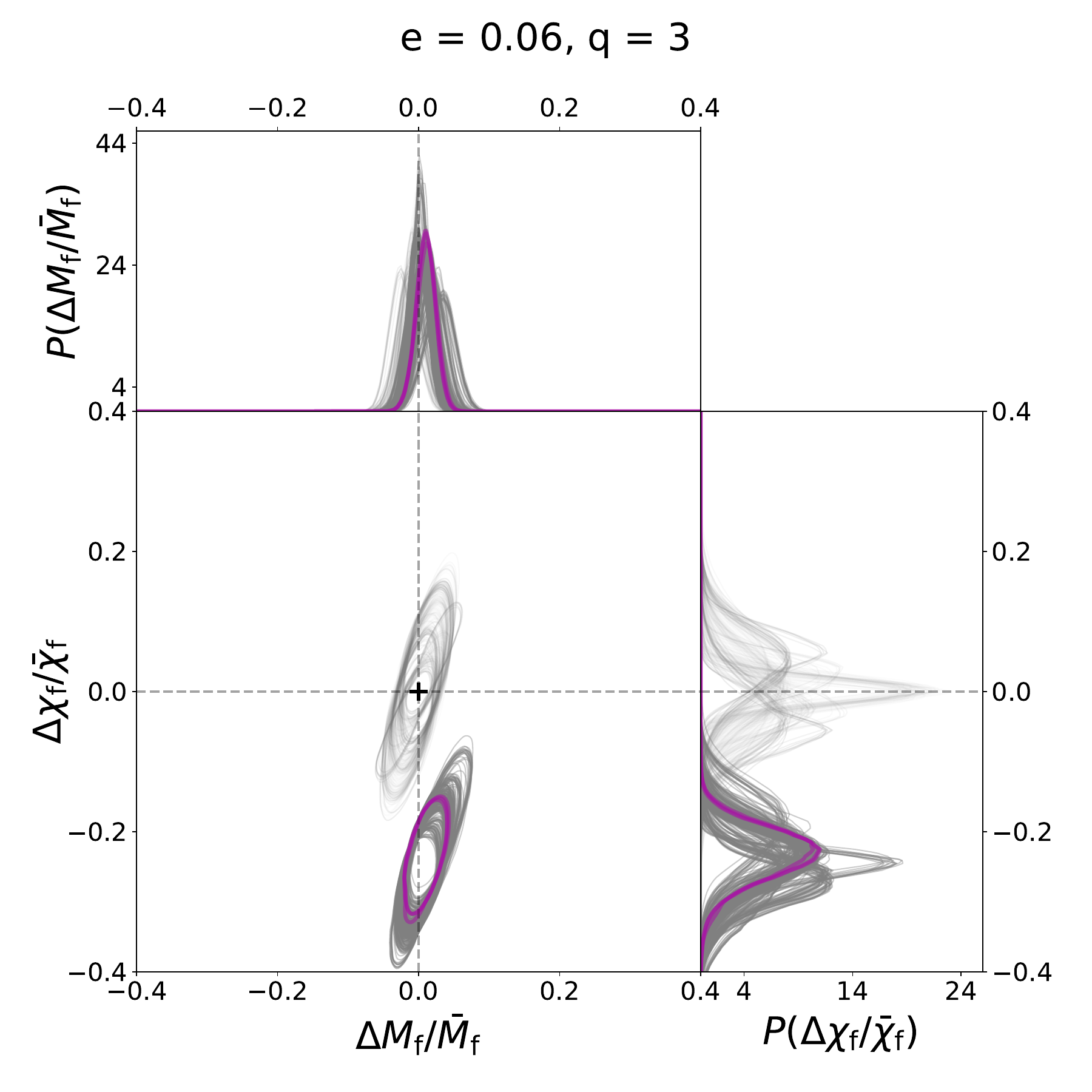}}
\qquad
\subfloat{\includegraphics[width=.5\textwidth]{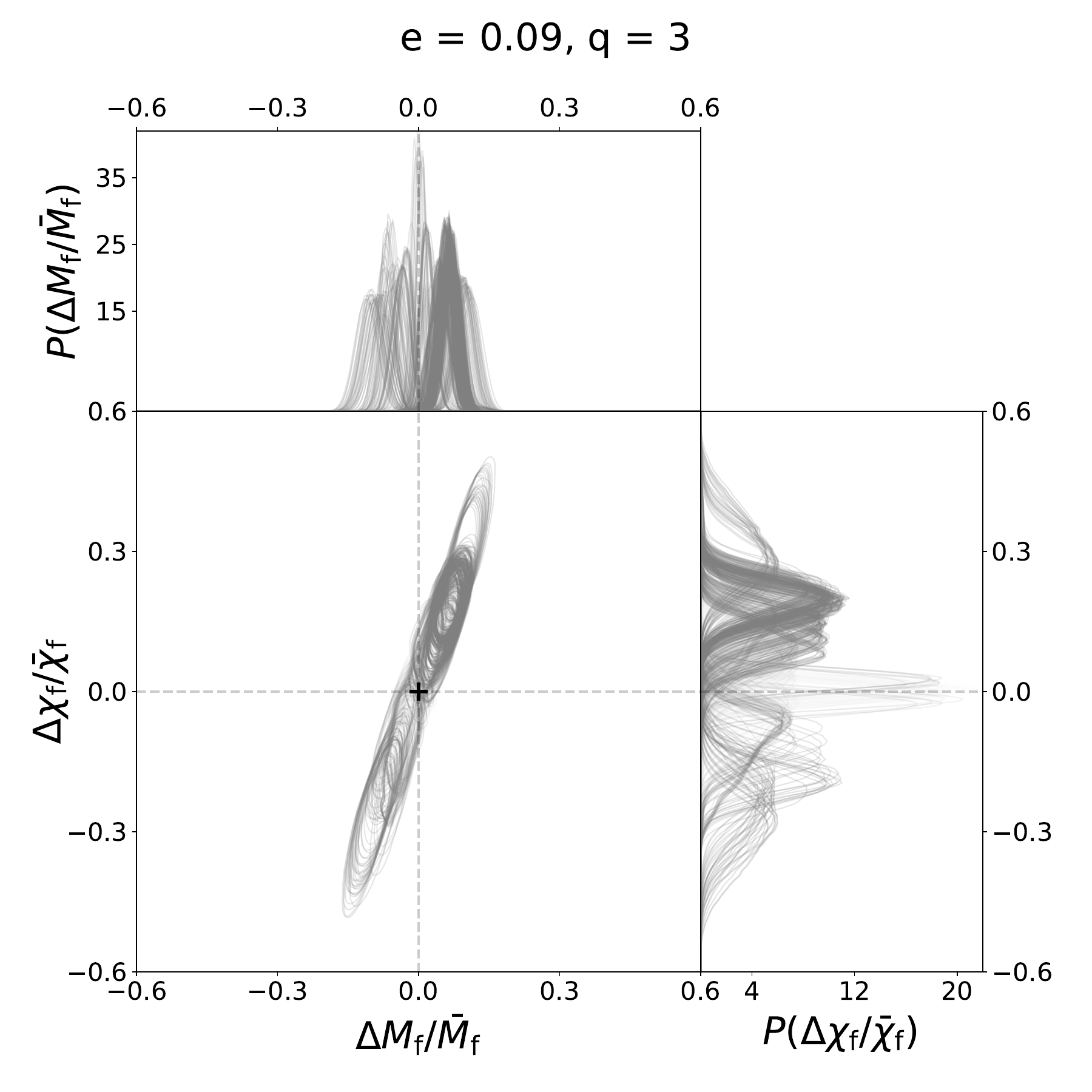}}

\caption[]{\label{fig:figure 5.6}Similar to figure \ref{fig:figure 5.4} for the $q = 3$ numerical relativity quasicircular and eccentric simulated observations.}

\end{figure}

We show the results obtained by performing meta IMRCT on the TIGER, FTI, and MDR test results applied on eccentric and quasicircular simulated signals from~\cite{Narayan:2023vhm} in table~\ref{tab:Table 5.3} and figures~\ref{fig:figure 5.4},~\ref{fig:figure 5.5} and~\ref{fig:figure 5.6}\; \footnote{The ordering of the pairs in these figures is the same as in the previous figures, except that since we now have the full set of testing parameters, each set of testing parameters (i.e.\ the $\{\delta\varphi_k\}$,$\{\delta\alpha_k\}$, $\{\delta\beta_k\}$ and $\{A_\alpha^{\gtrless 0}\}$) is ordered by increasing values of its index. Explicitly, for the MDR parameters, we have  $A_0^{<0}$, $A_0^{>0}$, $A_{0.5}^{<0}$, $A_{0.5}^{>0}$, $\ldots$}. The meta IMRCT finds that all quasicircular cases are consistent with GR at the 90\% credible level, with largest GR quantiles of $86.4\%$, $40.3\%$ and $37.4\%$ for $q = 1$, $2$ and $3$, respectively. This result is as expected since the tests use waveform models that describe binaries on quasicircular orbits, though the significantly larger maximum and median GR quantiles (and smaller Simes combined $p$-value) for the $q = 1$ simulated observation are unexpected: One expects that the waveform models used are all quite accurate for nonspinning quasicircular binaries with $q \leq 3$, and the GR PE finds only the expected small biases from the true values for the quasicircular simulated observations. The largest GR quantiles for $q = 1$ are for pairs with $A_{0.5}^{<0}$, and the $46$ largest GR quantiles (with values $\geq 55.6\%$) are for pairs with either $A_{0.5}^{<0}$ or $A_{0.5}^{>0}$. Both the $A_{0.5}$ analyses recover precessing binaries, and the posterior distributions of final masses and spins extend to larger values than the distribution from GR PE, particularly for the $A_{0.5}^{<0}$ analysis. However, the $A_{0.5}^{<0}$ analysis finds a notably smaller maximum likelihood than GR PE or the $A_{0.5}^{>0}$ analysis, so it may have been somewhat undersampled. This shows how the meta IMRCT can identify possible problems with analyses. However, when excluding the pairs with $A_{0.5}^{\gtrless 0}$, one still obtains a median GR quantile of $22.6\%$, considerably larger than the medians for the other quasicircular cases, though when excluding all the pairs with MDR, one obtains a median and maximum GR quantile of $17.0\%$ and $43.6\%$, and when also excluding the pairs with FTI, $11.5\%$ and $40.2\%$, showing that these are the analyses that give the largest bias away from GR in the meta IMRCT. 

    \begin{table}
      \centering
      \caption[Summary of meta IMRCT applied on numerically simulated quasicircular and eccentric signals]{\label{tab:Table 5.3}Summary of meta IMRCT applied on numerical relativity quasicircular and eccentric simulated observations. Here $q$ is the mass ratio and $e$ is the eccentricity at $\sim 17$~Hz. The TIGER, FTI and MDR tests are performed with all deviation parameters. The other columns have the same meanings as in table~\ref{tab:GW150914-like}.}
      \scalebox{0.66}{
      \begin{tabular}{*{25}{c}}
	\hline\hline
	\\
        \multirow{2}{*}{Simulation} & \multirow{2}{*}{$q$ }& \multirow{2}{*}{$e$}  & \multirow{2}{*}{GR PE}& \multirow{2}{*}{IMRCT}& \multirow{2}{*}{TIGER} &\multirow{2}{*}{FTI }& \multirow{2}{*}{MDR}& \multirow{2}{*}{GR Quantile (\%)}  & \multirow{2}{*}{Gaussian $\sigma$} & Combined $p$-value& \multirow{2}{*}{$N^{100\%}/N_\text{mIMRCT}^{\text{grtr}, 90\%}/N_\text{mIMRCT}^\text{grtr}/N_\text{tot}$} \\
              &   &  & & & &  & && &$p_\text{S}, p_\text{H}$&\\
        
       \\
        \hline
        
        SXS:BBH:1155 & $1$ & $0.0$ & \checkmark& $\times$& All & All & All & $25.4^{+34.4}_{-14.5}$ &  $0.3^{+0.5}_{-0.2}$ & $0.907, 1$ &$0/0/240/703$ \\
        \\
        SXS:BBH:1355 & $1$ & $0.05$ & \checkmark& $\times$& All & All & All & $37.0^{+48.9}_{-32.5}$ & $0.5^{+1.0}_{-0.4}$ & $0.963, 1$ &$0/0/2/703$\\ 
        \\
        SXS:BBH:1357 & $1$ & $0.10$ & \checkmark& $\times$& All & All & All& $99.2^{+0.8}_{-86.7}$ & $2.6^{+0.4}_{-2.5}$ & $6.98\times 10^{-3}, 4.98\times 10^{-2}$ &$297/15/18/703$ \\
        \\
        SXS:BBH:1222 & $2$ & $0.0$ & \checkmark& $\times$& All & All & All & $6.1^{+20.7}_{-2.6}$ &  $0.1^{+0.3}_{-0.0}$ & $0.972, 1$ &$0/0/31/703 $\\
        \\
        SXS:BBH:1364 & $2$ & $0.05$ & \checkmark& $\times$& All & All & All & $46.3^{+45.3}_{-41.7}$ &  $0.6^{+1.1}_{-0.6}$ & $0.848, 1$ &$0/0/41/703$\\
        \\
        SXS:BBH:1368 & $2$ & $0.10$ & \checkmark& $\times$& All & All & All & $87.9^{+12.1}_{-77.6}$ & $1.6^{+1.4}_{-1.4}$ & $3.15\times 10^{-2}, 0.225$ &$43/107/126/703$ \\
        \\
        SXS:BBH:2265 & $3$ & $0.0$ & \checkmark& $\times$& All & All & All & $4.9^{+16.5}_{-2.6}$ & $0.1^{+0.2}_{-0.0}$ &$0.981, 1$ &$0/0/33/703 $\\
        \\
        SXS:BBH:1371 & $3$ & $0.06$ & \checkmark& $\times$& All & All & All & $44.5^{+55.5}_{-42.2}$ &  $0.6^{+2.4}_{-0.6}$ &$6.28\times 10^{-3}, 4.48\times 10^{-2}$ &$112/224/225/703$\\
        \\
        SXS:BBH:1373 & $3$ & $0.09$ & \checkmark& $\times$& All & All & All & $96.9^{+3.1}_{-88.2}$ & $2.2^{+0.8}_{-2.0}$ & $7.48\times 10^{-3}, 5.34\times 10^{-2}$ &$282/0/11/703$\\ 
            
        \hline\hline
        \end{tabular}}
        \end{table}

\newpage

Also as expected, the median and maximum meta IMRCT GR quantiles increase with eccentricity (the maxima are $99.5\%$, $95.4\%$ and $100\%$ for $q = 1$, $2$ and $3$ with the smaller eccentricity and $100\%$ for all three mass ratios with the larger eccentricity), with the median value excluding GR at $>2\sigma$ Gaussian equivalent for the $q = 1$ and $3$ larger eccentricity simulated observations. However, the Simes combined $p$-values only decrease (i.e.\ become more significant) monotonically with increasing eccentricity for $q$ = 2. Specifically, for $q = 1$, the quasicircular Simes $p$-value is slightly smaller than the one for the smaller eccentricity, while for $q = 3$, both the Simes and Hommel $p$-values for the smaller eccentricity are a bit below the ones for the larger eccentricity, and in fact give the smallest values found for this set of simulated observations. However, the combined $p$-values for the cases with the larger eccentricity are always fairly significant, ranging from $3.15\times 10^{-2}$ for $q = 2$ to $6.98\times10^{-3}$ for $q = 1$. The meta IMRCT also gives a GR quantile that is larger than $90\%$ and larger than that from either of the individual tests for the $q = 1$ and $2$ simulated observations with the larger eccentricities, and the $q = 3$ simulated observation with the smaller eccentricity. For the $q = 1$ simulated observation with the larger eccentricity, the $15$ meta IMRCT GR quantiles in such cases range from $98.9\%$ to $100\%$ and all of the pairs involve $\delta\varphi_k^\text{FTI}$, $k \in \{6, 7\}$ paired with either GR PE or MDR. The largest difference between the meta IMRCT GR quantile and the largest individual GR quantile in these cases is $1.1$~percentage points for the $(\delta\varphi_7^\text{FTI}, A_{1.5}^{<0})$ and $(\delta\varphi_7^\text{FTI}, A_{2.5}^{>0})$, both of which have a meta IMRCT GR quantile of $100\%$; the $\delta\varphi_7^\text{FTI}$, $A_{1.5}^{<0}$ and $A_{2.5}^{>0}$ GR quantiles are $98.8\%$, $24.4\%$ and $93.1\%$, respectively. There are also $297$ pairs ($42\%$ of the total) where the meta IMRCT gives a GR quantile of $100\%$ and at least one of the individual tests does, as well.

For the $q = 2$ simulated observation with the larger eccentricity, the $107$ cases ($15\%$ of the total) with meta IMRCT GR quantile $> 90\%$ and greater than the individual GR quantiles are mostly FTI--MDR pairs, as well as some GR PE--FTA pairs and the $\delta\varphi_7^\text{TIGER}$ and $\delta\beta_2$ pairs with $A_{2.5}^{>0}$. The largest meta IMRCT GR quantiles in this case are $100\%$, and they can be significantly larger than the individual GR quantiles, in particular being greater by $42.6$ percentage points in $9$ cases with $A_{2.5}^{>0}$ in the pair, including its pair with $A_{2.5}^{<0}$. The largest difference between the meta IMRCT GR quantile and the maximum individual GR quantile is for the pair $(\text{GR PE}, \delta\varphi_3^\text{FTI})$, where meta IMRCT gives a GR quantile of $90.4\%$ and $\delta\varphi_3^\text{FTI}$ gives a GR quantile of $4.8\%$.

For the $q = 3$ simulated observation with the larger eccentricity, there are $282$ pairs ($40\%$ of the total) with a meta IMRCT GR quantile of $100\%$ (where at least one of the individual tests does, as well) and $391$ pairs ($56\%$ of the total) with a meta IMRCT GR quantile $> 90\%$, though there are no cases where the meta IMRCT gives a larger GR quantile than the individual tests do. The pairs that give a GR quantile of $100\%$ are mostly pairs of MDR with another analysis, but also involve pairs of TIGER and FTI with GR PE, and TIGER and FTI inspiral with $\delta\alpha_k$, as well as $(\delta\beta_3, \delta\alpha_2)$.

However, for the $q = 3$ simulated observation with the smaller eccentricity, there are $224$ cases ($32\%$ of the total) with meta IMRCT GR quantile $> 90\%$ and greater than the individual GR quantiles. All of these have a meta IMRCT GR quantile of $100\%$ and consist of pairs of GR PE, TIGER, or FTI with MDR. The largest difference between the meta IMRCT GR quantile and the largest individual GR quantile is $43.0$ percentage points for the $6$ pairs of GR PE, $\delta\beta_3$ or $\delta\alpha_4$ with $A_{3}^{\gtrless 0})$, where the meta IMRCT gives a GR quantile of $100\%$, while $\delta\beta_3$, $\delta\alpha_4$ and $A_3$ have GR quantiles of $17.9\%$, $20.1\%$ and $57.0\%$, respectively. There are also $112$ cases where the meta IMRCT and at least one of the individual tests both give a GR quantile of $100\%$.

For the $q = 1$ simulated observation with the smaller eccentricity, the largest meta IMRCT GR quantile is $99.5\%$, for the $(\delta\beta_3, \delta\alpha_3)$ pair, where the individual GR quantiles are $100\%$ and $98.9\%$, respectively. For the $q = 2$ simulated observation with the smaller eccentricity, the largest meta IMRCT GR quantile is $95.4\%$, for the $(\delta\varphi_1^\text{TIGER}, A_3^{>0})$ pair, where the individual GR quantiles are $98.8\%$ and $88.0\%$, respectively.

\subsubsection{Choice of parameters for meta IMRCT}
\label{sssec:luminosity-distance}

As mentioned above, one could choose other GR parameters (intrinsic or extrinsic) to check the consistency between the results of the two analyses in the meta IMRCT. Here we briefly explore if the luminosity distance could be a better parameter instead of the final mass and spin, particularly since in the MDR test, $A_\alpha$ correlates with luminosity distance and hence the posterior on luminosity distance in the MDR analysis might be different from that in other analyses. For this check, we chose the simulated observation that gives the maximum GR quantiles for MDR and GR PE analysis pairs. We found this to be the $q=2$ lower eccentricity case (SXS:BBH:1371), which gives GR quantiles of $100\%$. In fact, the $90\%$ credible intervals on the final spin are disjoint for all these pairs, though the final mass intervals overlap. We thus checked how well the posteriors on the luminosity distance agree for these pairs and found considerable overlap of the posteriors, where the median of the GR PE luminosity distance posterior lies within the $90\%$ bound for all the MDR luminosity distance posteriors. Thus, the luminosity distance would not be a good parameter to use to detect this sort of deviation from the model used in performing the meta IMRCT.

\begin{table*}[h!]
    \caption[Summary of meta IMRCT on real events]{\label{tab:real_events} Summary of meta IMRCT on real events, similar to table~\ref{tab:GW150914-like}. }
    \centering
    \scalebox{0.7}{
    \begin{tabular}{c c c c c c c c c c}
         \hline \hline \\
        \multirow{2}{*}{Event} & \multirow{2}{*}{GR PE}& \multirow{2}{*}{IMRCT} &\multirow{2}{*}{TIGER} & \multirow{2}{*}{FTI} & \multirow{2}{*}{MDR}& \multirow{2}{*}{GR Quantile (\%)}   & \multirow{2}{*}{Gaussian $\sigma$}  & Combined $p$-value & \multirow{2}{*} {$N^{100\%}/N_\text{mIMRCT}^{\text{grtr}, 90\%}/N_\text{mIMRCT}^\text{grtr}/N_\text{tot}$ }
        \\ 
        &   &  & & & &  & &$p_\text{S}, p_\text{H}$&\\
        &   &  & & & &  & & &
        \\
         \hline
        GW170817  & \checkmark&$\times$ & All & All &  $\times$ &$4.2^{+14.3}_{-3.3}$ & $0.0^{+0.2}_{-0.0}$ & $0.997, 1$ &$0/0/0/210$ \\
  \\
    GW190412  & \checkmark&I, MR & All & All & All &  $14.3^{+38.2}_{-10.9}$ & $0.2^{+0.5}_{-0.1}$ & $0.998, 1$ &$0/0/38/945$ \\
    \\
  GW190521  & \checkmark& $\times$ & All &$\times$ & $\times$ & $4.2^{+4.7}_{-1.2}$ & $0.0^{+0.0}_{-0.0}$ & $0.971, 1$ &$0/0/0/15$ \\
 \\
  GW190814 & \checkmark& $\times$ & All & All & All & $6.2^{+40.0}_{-5.8}$ & $0.1^{+0.5}_{-0.1}$ & $0.999, 1$ &$0/0/19/861$ \\
\\
  GW200225\_060421  & \checkmark & I, MR & $\times$ & All & All & $59.1^{+14.1}_{-52.0}$ & $0.8^{+0.3}_{-0.7}$ & $0.694, 1$ &$0/0/2/405$ \\
  
  \hline\hline
    \end{tabular} }
\end{table*}

\begin{figure}[htb]
    \centering

    \subfloat{\includegraphics[width=.5\textwidth]{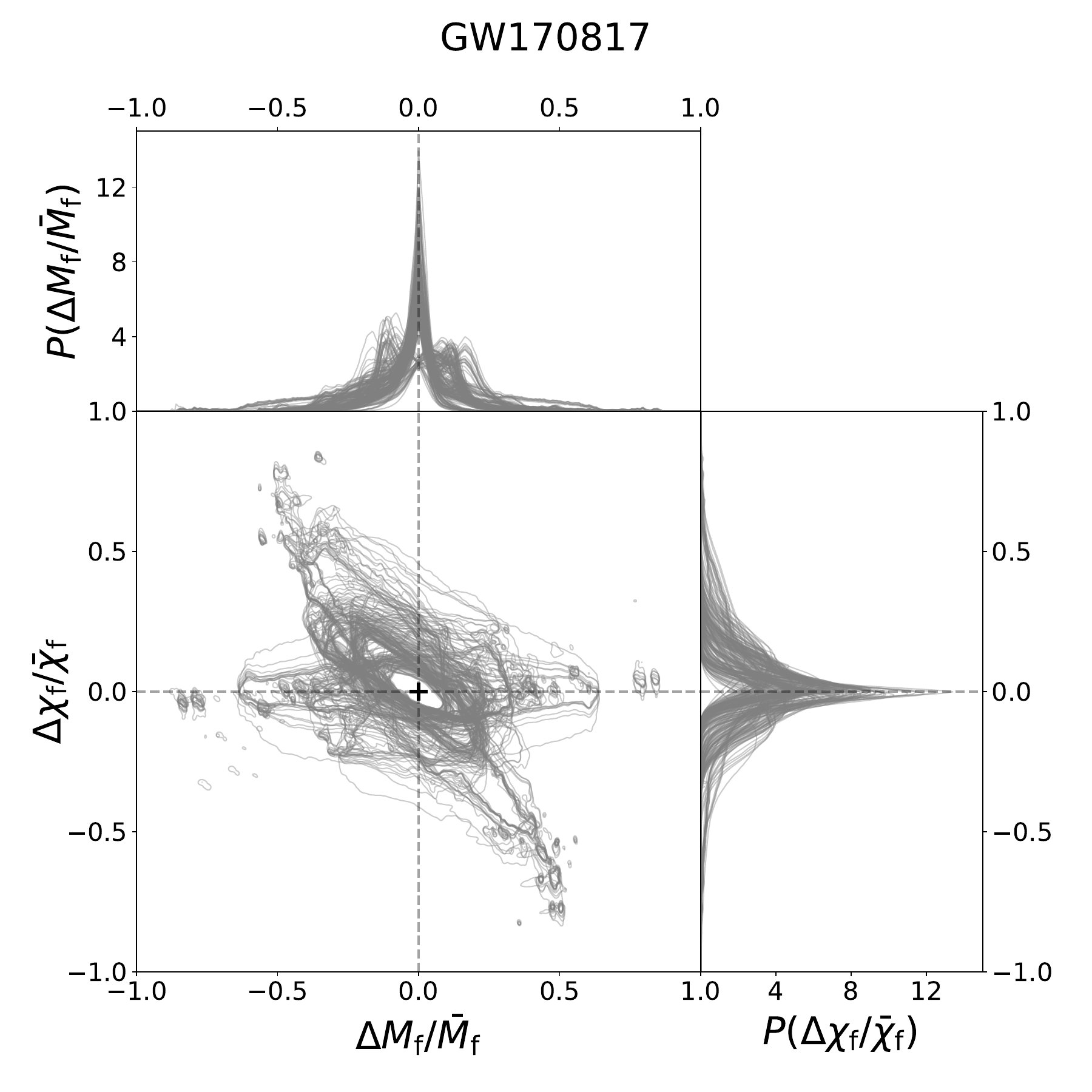}}
\qquad
\subfloat{\includegraphics[width=.5\textwidth]{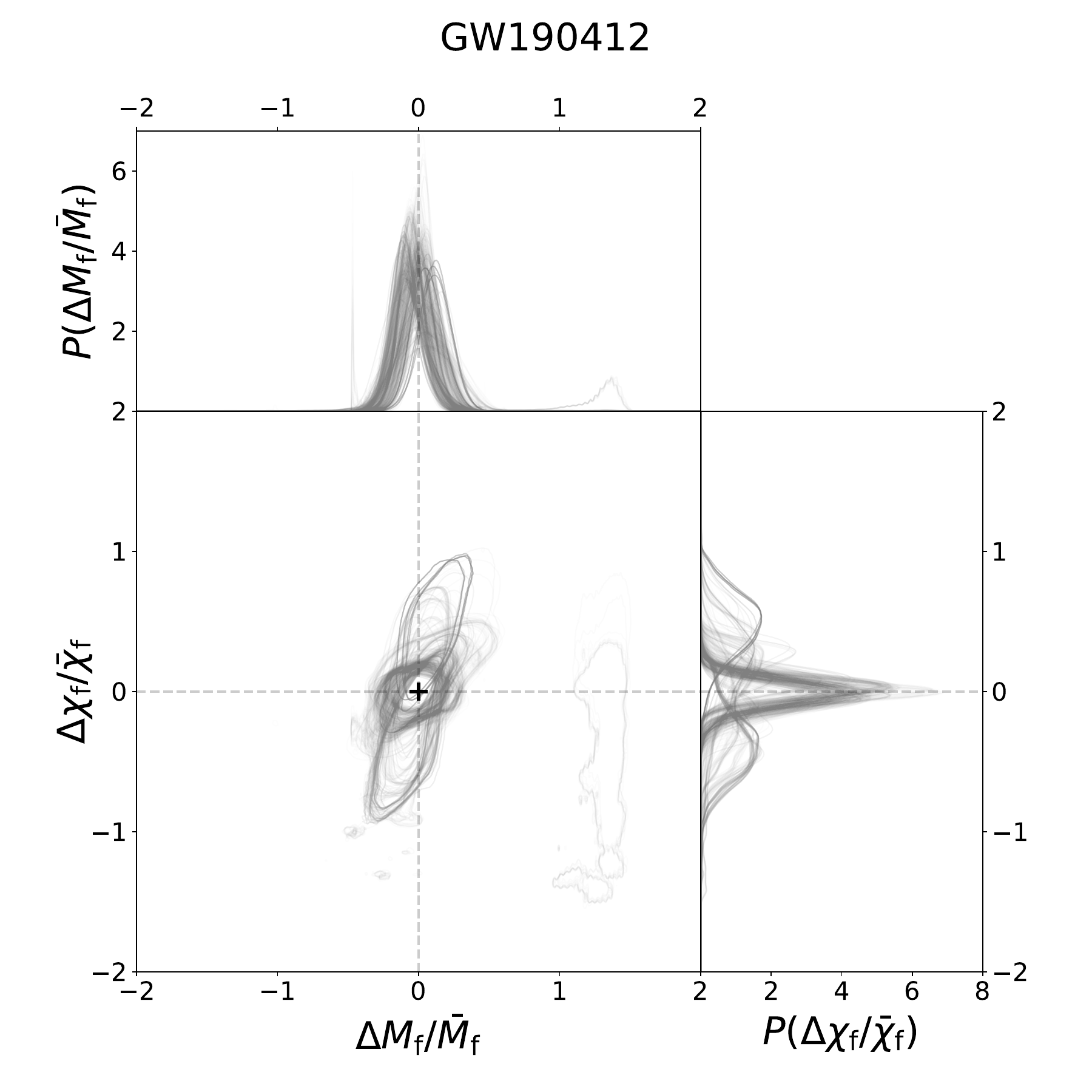}}
\\
    \subfloat{\includegraphics[width=.5\textwidth]{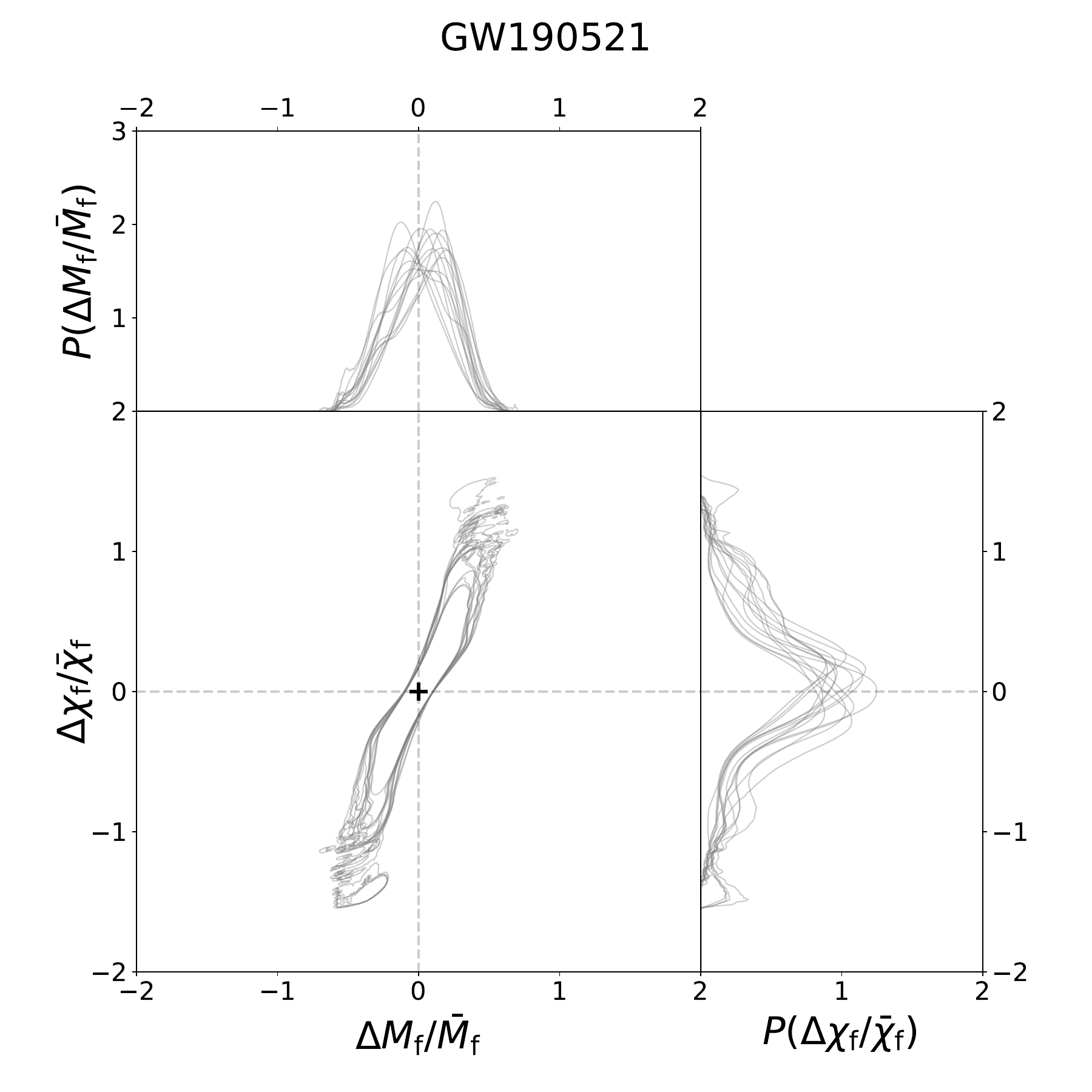}}
\qquad
\subfloat{\includegraphics[width=.5\textwidth]{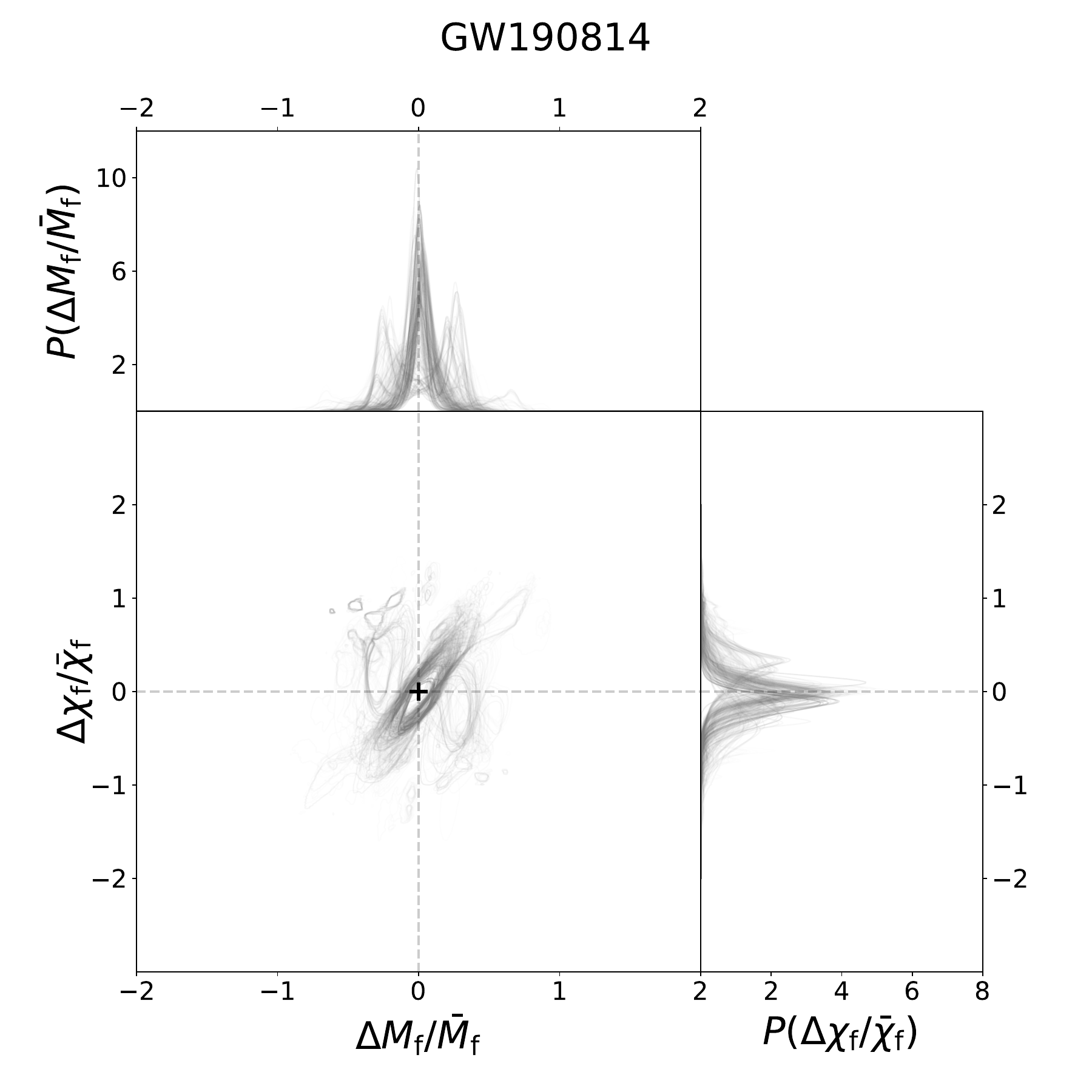}}
\\
\subfloat{\includegraphics[width=.5\textwidth]{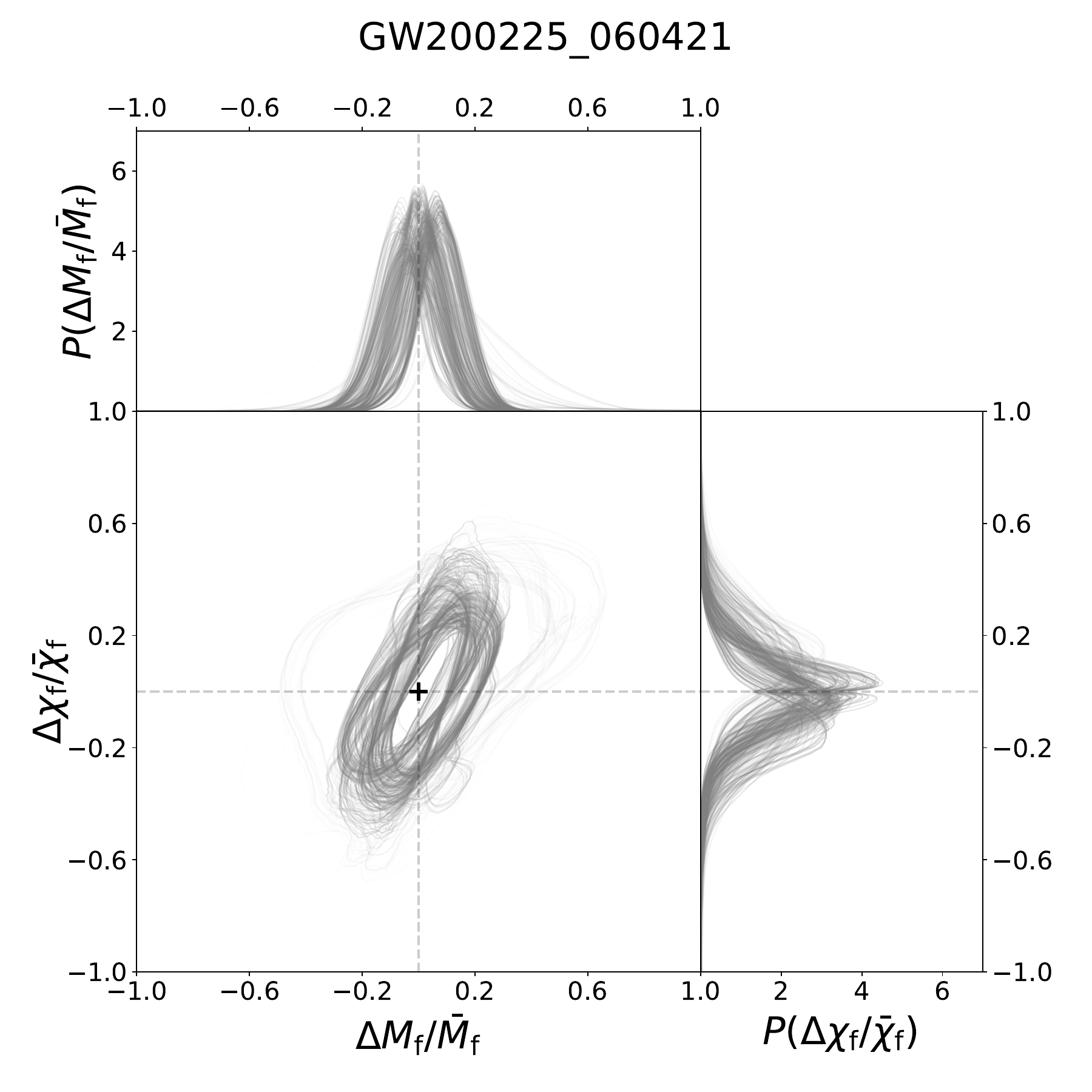}}

\caption[]{\label{fig:real_events}Result of meta IMRCT on real events presented as 2d contour plots and plotted in the same way as in figure~\ref{fig:GW150914-like_larger}.}

\end{figure}

\subsection{Real events}
\label{ssec:real_events}

We now consider five real events: four exceptional events and one event that showed problematic results in the LVK O3b testing GR analyses~\cite{O3b_TGR}. The exceptional events are the likely binary neutron star GW170817~\cite{GW170817} (see also~\cite{GWTC-1_paper}); the unequal-mass binary GW190412~\cite{GW190412}; the high-mass event GW190521~\cite{GW190521,GW190521_prop}; and the very unequal-mass event GW190814~\cite{GW190814}, whose secondary is likely a light black hole, but could also be a heavy neutron star or something more exotic. See also~\cite{GWTC-2.1_paper} for the LVK's latest analyses of the latter three signals. The problematic event is GW200225\_060421~\cite{GWTC-3_paper}, which was an outlier in the LVK MDR results and had the smallest $p$-value in the residuals test ($0.05$) of all the O3 events~\cite{O3b_TGR}. It also had a larger significance in the minimally modelled coherent WaveBurst search~\cite{Klimenko:2015ypf} than in the templated searches that use GR compact binary waveforms~\cite{GWTC-3_paper}.

We apply the meta IMRCT to these events using the LVC/LVK GR PE results and their testing GR results for the tests we considered in previous sections (also including the $-1$PN test, with testing parameter $\delta\varphi_{-2}$, for both FTI and TIGER) from~\cite{GW170817_TGR,O3a_TGR,O3b_TGR}, using the data from~\cite{GW170817_TGR_data_release,O3a_TGR_data_release,O3b_TGR_data_release}. The waveform models employed for GR parameter estimation of these events is carried out using IMRPhenomPv2\_NRTidal (GW170817), IMRPhenomPv3HM (GW190412, GW190521 and GW190814) and IMRPhenomXPHM (GW200225\_060421). The tests we consider are based on these waveforms except for FTI, which uses SEOBNRv4HM\_ROM, and MDR, which uses IMRPhenomPv2 for GW190412, GW190521 and GW190814. The TIGER and FTI tests for GW170817 use IMRPhenomPv2\_NRTidal and SEOBNRv4\_ROM\_NRTidal, respectively. We show the meta IMRCT results in table~\ref{tab:real_events} and figure~\ref{fig:real_events}, finding consistency with GR, as in the LVK results for the non-problematic events \footnote{While GW170817 is likely a binary neutron star, and the tests of GR are applied using waveforms that include tidal effects~\cite{Dietrich:2018uni}, in our application of the meta IMRCT we use the BBH final mass and spin fits, so we only consider the consistency of the initial masses and spins, not the tidal deformabilities.}. Specifically, the maximum GR quantile we obtain is $89.4\%$ for the $(\delta\varphi_{-2}^\text{FTI}, \delta\alpha_4)$ pair and GW190412; the specifics of the maxima for each event are given in table~\ref{tab:real_events_max_GR_quant}. While the meta IMRCT does not find any GR quantiles $> 90\%$ for the problematic event GW200225\_060421, the median GR quantile of $59.1\%$ is considerably larger than for the other real events or GR simulated observations. We also found that the individual FTI GR quantiles for this event are $> 90\%$ for the $1.5$PN ($\delta\varphi_{3}$) and higher testing parameters, indeed as large as $94.4\%$ for the $2.5$PN log ($\delta\varphi_{5l}$) testing parameter, though this is not remarked on in the LVK testing GR paper, which only presents upper bounds for individual events. The combined $p$-values are also very close to $1$, with the smallest (i.e.\ most significant GR deviation) for GW200225\_060421, though at $0.694$ it is not at all statistically significant.

To assess the potential effect of a non-uniform distribution of the GR quantile for the meta IMRCT in Gaussian noise, we take the distribution of GR quantiles obtained for the standard IMRCT applied to a population of simulated observations in Gaussian noise as the fiducial one. The p-p plot for this distribution is shown in the left panel of figure~\ref{fig:pp} and is not very inconsistent with a uniform distribution, but parts of the p-p plot are outside of the $95\%$ confidence band. We thus compute the median and $90\%$ interval for the GR quantiles and Simes combined $p$-value after first applying the interpolated CDF of the GR quantiles from the meta IMRCT in Gaussian noise to the GR quantiles for the real events. That is, we are performing the probability integral transform, so that the resulting distribution of GR quantiles would be uniform if they followed the distribution whose CDF we are using. In all cases, the differences in the medians are at most $5.8$ percentage points and the differences in the $90\%$ bounds of the GR quantiles are at most $10.9$ percentage points. The Simes combined $p$-value becomes $1$, except for GW200225\_060421, where it becomes $0.566$. Thus, we see that while there are some nonnegligible differences in these results if one assumes that the distribution of the GR quantile for the meta IMRCT in Gaussian noise is the true distribution of GR quantiles for the meta IMRCT, these are not significant enough to change the qualitative conclusions.

\begin{table}[h!]
    \caption[]{\label{tab:real_events_max_GR_quant} The details of the meta IMRCT pairs that give the largest GR quantile for the real events.}
    \centering
    \begin{tabular}{c c c c }
         \hline \hline \\
        \multirow{2}{*}{Event} & Max.\ mIMRCT & \multirow{2}{*}{Pair} & \multirow{2}{*}{Indiv.\ GR quantiles (\%)}\\
        & GR Quantile (\%)  & & \\
        \\
         \hline
        GW170817 & $45.1$ & $(\delta\varphi_0^\text{TIGER}, \delta\varphi_4^\text{TIGER})$ & $81.1, 82.2$\\
	\\
	GW190412 & $89.4$ & $(\delta\varphi_{-2}^\text{FTI}, \delta\alpha_4)$ & $65.0, \hphantom{0}0.5$\\
	\\
	GW190521 & $11.4$ & $(\delta\alpha_2, \delta\alpha_4)$ & $10.3, 49.5$\\
 	\\
	GW190814 & $77.2$ & $(\delta\varphi_{-2}^\text{TIGER}, \delta\beta_3)$ & $94.5, 78.1$ \\
	\\
	GW200225\_060421 & $79.0$ & $(\delta\varphi_6^\text{FTI}, A_{1.5}^{<0})$ & $94.3, 95.9$\\
  \hline\hline
    \end{tabular}
\end{table}

Due to the specific mass priors used in the different tests for GW190412, GW190814 and GW200225\_060421, the priors on the deviation parameters peak well away from the GR value for some pairs (distances of $> 0.4$ from the GR value). The cases that peak the furthest away from the GR value are any of the pairs of GR PE, FTI or MDR with the IMR consistency inspiral or merger-ringdown analyses for GW190412 and with TIGER for GW190814. In all these cases, the peak of the posterior is around $(\Delta M_f/\bar{M}_f, \Delta \chi_f/\bar{\chi}_f) = (1.3, 0)$, with the ordering of the pairs we use, except for the GR PE pair, where it has the opposite sign. Nevertheless, in all cases the posteriors still peak around the GR value. We find that for GW190814 the significant shift of the prior away from the region preferred by the posterior causes some artefacts when reweighting certain pairs to a flat prior in $(\Delta M_f/\bar{M}_f, \Delta \chi_f/\bar{\chi}_f)$. We are able to mitigate the artefacts by imposing a threshold on the probability density of $8 \times 10^{-4}$ of the maximum, which only changes the GR quantiles of the pairs without artefacts by less than $1.6$ percentage points. The largest GR quantile difference are for pairs with FTI. The largest difference for non-FTI pairs is $0.8$ percentage points for the $(\delta\varphi_{-2}^\text{TIGER}, \delta\beta_2)$ pair.

For GW190412, we find that for all the aforementioned pairs where the prior peaks well away from the GR value the reweighted posteriors are all centred around zero, with no secondary peaks, while the unreweighted posteriors have a secondary peak around the peak of the prior. However, for the pairs of the TIGER analyses with the merger-ringdown analysis, where the prior peaks around $(\Delta M_f/\bar{M}_f, \Delta \chi_f/\bar{\chi}_f) = (0.8, 0)$, the unreweighted posterior still peaks at about the same location as for the FTI and merger-ringdown pairs, so one obtains the secondary peaks in the reweighted posterior seen in figure~\ref{fig:real_events}.

\section{Conclusions}
\label{sec:concl}
Gravitational waves from coalescing compact binaries are now routinely used to test GR in highly dynamical, non-linear gravity regions of space-time. There are now quite a few tests that have been applied to inspect the consistency of the LVK observations with GR predictions but no deviations from GR have been reported so far. In this paper, we proposed a new test of GR, called the meta IMRCT, that checks the consistency between the results from two different tests of GR analyses (or a GR analysis) on a given compact binary coalescence. Specifically, the meta IMRCT checks the consistency of the inference of the final mass and spin of the remnant black hole from two different tests of GR analyses or the standard GR analysis of the signal. 

The meta IMRCT is a generalized version of the standard IMRCT (proposed in~\cite{Ghosh:2016qgn,Ghosh:2017gfp}) that instead checks the consistency between the final mass and spin of the remnant estimated from independently analyzing the low- and high-frequency portions of the signal. Unlike the standard IMRCT, which is restricted to binaries whose masses are neither too low nor too high, so that there is sufficient SNR in both the inspiral and merger-ringdown portions of the signal (see, e.g.\ \cite{O3b_TGR}), the meta IMRCT can be applied to BBHs of any total mass and SNR. It can even be applied to binaries involving neutron stars. Here one can use the BBH final mass and spin fits since these fits are just used to conveniently map the initial mass and spin parameter space to a two-dimensional space. However, it would also be possible to use fits for the final mass and spin of binary neutron star or neutron star--black hole binaries (e.g.\ the ones in \cite{Deng:2020rnf} and \cite{Zappa:2019ntl}; the latter are just for neutron star--black hole binaries, but are more accurate, though the accuracy is not a great concern here, since these fits are just used to provide a convenient, physically inspired dimensionality reduction). The meta IMRCT is also computationally inexpensive to perform, without any additional stochastic sampling or evaluation of waveforms, once one has obtained the individual test results.

To assess the performance of the meta IMRCT we applied it to a variety of simulated GR and non-GR signals from quasicircular BBHs from~\cite{Johnson-McDaniel:2021yge} and numerical relativity GR signals from quasicircular and eccentric BBHs from~\cite{Narayan:2023vhm} (see section~\ref{sec:sim_obs} for details). Specifically, we utilize the tests of GR and standard GR parameter estimation analyses (all using quasicircular waveform models) performed on these simulated signals in~\cite{Johnson-McDaniel:2021yge,Narayan:2023vhm} and compare the final mass and spin estimates from these in pairs. The tests of GR that are performed in~\cite{Johnson-McDaniel:2021yge,Narayan:2023vhm} are the two parameterized tests, TIGER and FTI, the modified dispersion relation test, and the standard IMRCT (see section~\ref{sec:tests} for detail). We apply the meta IMRCT to all the different testing parameters for TIGER, FTI, and the modified dispersion relation test and consider the inspiral and merger-ringdown analyses of the standard IMRCT separately. (Of course, the meta IMRCT reproduces the standard IMRCT results when applied to the inspiral and merger-ringdown analyses.)

We found that the meta IMRCT indeed finds all simulated GR quasicircular signals (Phenom, EOB, or numerical relativity) are consistent with GR, as expected since the waveform models used in all the individual tests are quasicircular. The meta IMRCT reports high GR quantiles (meaning large GR deviations) for signals with larger GR deviations, larger than the individual tests' GR quantiles in some cases. For simulations with smaller GR deviation, the meta IMRCT does not report high GR quantiles, similar to the individual test analyses. The combined $p$-values also are small (so high significance for the GR deviation) for the large GR deviation and larger for the smaller GR deviation.  For numerical relativity eccentric simulations the meta IMRCT GR quantiles increase with eccentricity, as expected, and give significantly large GR quantities ($\sim 100\%$) for larger eccentricity simulations. For one eccentric case, the meta IMRCT finds a GR deviation for several pairs where the individual tests to which it is being applied do not find a significant deviation, and finds a GR quantile $> 90\%$ and larger than the individual GR quantiles for almost a third of the pairs. This case also gives the smallest combined $p$-value.

We also applied the meta IMRCT to real compact binary signals GW170817, GW190412, GW190521, GW190814 and GW200225\_060421 using the results from the LV(K) collaboration's application of the tests of GR (we considered only TIGER, FTI, MDR and IMRCT). The meta IMRCT finds all these events consistent with GR, giving GR quantiles below $90\%$ and combined $p$-values close to $1$.

Similar to other tests of GR, the meta IMRCT could also produce false GR violations due to various causes, as discussed in great detail in~\cite{Gupta:2024gun}. One of the primary causes of false GR violations in the meta IMRCT is waveform model inaccuracies and missing physics. If the waveform model used in either of the two tests of GR or standard GR analyses used in the meta IMRCT is insufficiently accurate, this can lead to a false deviation from GR. Moreover, even if the waveform models used are sufficiently accurate descriptions of the waveform from a quasicircular binary in GR, we have already shown that the meta IMRCT will give an apparent GR deviation if it is applied to a signal from a binary with sufficiently large eccentricity. We would also expect the meta IMRCT to return apparent deviations from GR in other cases where the signal includes physics not included in the waveform model used, notably strong lensing or environmental effects, where studies have found deviations from GR in individual tests (see~\cite{Narayan:2024rat} and~\cite{Roy:2024rhe}, respectively).


In the future, it will be interesting to explore applying the meta IMRCT using other GR parameters (such as different combinations of the masses and spins), or larger numbers of parameters, to see what gives the best performance. It will also be interesting to study the performance of the meta IMRCT on simulated binary neutron star and neutron star--black hole binaries, in particular seeing whether including the tidal parameters improves the meta IMRCT.

\section*{Acknowledgments}
We thank Muhammed Saleem, N.~V.~Krishnendu, James Clark and Purnima Narayan for their work on producing the results from~\cite{Johnson-McDaniel:2021yge,Narayan:2023vhm} that we use. We also thank the anonymous referee for their valuable comments and suggestions, which have helped improve the presentation of this manuscript. NKJ-M is supported by NSF grant AST-2205920. AG is supported in part by NSF grants AST-2205920 and PHY-2308887.

This research has made use of data obtained from the Gravitational Wave Open Science Center (\url{https://www.gwosc.org/}), a service of the LIGO Laboratory, the LIGO Scientific Collaboration and the Virgo Collaboration. LIGO Laboratory and Advanced LIGO are funded by the United States National Science Foundation (NSF) as well as the Science and Technology Facilities Council (STFC) of the United Kingdom, the Max-Planck-Society (MPS), and the State of Niedersachsen/Germany for support of the construction of Advanced LIGO and construction and operation of the GEO600 detector. Additional support for Advanced LIGO was provided by the Australian Research Council. Virgo is funded, through the European Gravitational Observatory (EGO), by the French Centre National de Recherche Scientifique (CNRS), the Italian Istituto Nazionale di Fisica Nucleare (INFN) and the Dutch Nikhef, with contributions by institutions from Belgium, Germany, Greece, Hungary, Ireland, Japan, Monaco, Poland, Portugal, Spain. The construction and operation of KAGRA are funded by Ministry of Education, Culture, Sports, Science and Technology (MEXT), and Japan Society for the Promotion of Science (JSPS), National Research Foundation (NRF) and Ministry of Science and ICT (MSIT) in Korea, Academia Sinica (AS) and the Ministry of Science and Technology (MoST) in Taiwan.

This study used the software packages LALSuite~\cite{lalsuite},  Matplotlib~\cite{Hunter:2007ouj}, NumPy~\cite{Harris:2020xlr}, PESummary~\cite{Hoy:2020vys}, and SciPy~\cite{Virtanen:2019joe}. 
This is LIGO document P2400174.

\bigskip

\bibliographystyle{iopart-num}
\bibliography{meta_imrct}

\end{document}